\documentclass[aps,prb,twocolumn,notitlepage,amsfonts,citeautoscript,superscriptaddress,showpacs]{revtex4-1}

\usepackage{natbib}
\usepackage{listings}
\usepackage{color}

\definecolor{dkgreen}{rgb}{0,0.6,0}
\definecolor{gray}{rgb}{0.5,0.5,0.5}
\definecolor{mauve}{rgb}{0.58,0,0.82}

\usepackage{amsmath,amssymb,mathrsfs}
\usepackage{latexsym}
\usepackage{graphicx} 
\usepackage{grffile}
\usepackage{epstopdf}
\usepackage{amsfonts}
\usepackage{hyperref}
\usepackage{url}
\usepackage{bm}
\usepackage{times}

\usepackage{xr}

\graphicspath{{./}{fig/}{fig/fig_rrc/}}

\begin{document}
\title{Precursor of Laughlin state of hard core bosons on a two--leg ladder}

\author{Alexandru Petrescu}  
\affiliation{Department of Electrical Engineering, Princeton University, Princeton, New Jersey, 08544}

\author{Marie Piraud}
\affiliation{Department of Physics and Arnold Sommerfeld Center for Theoretical Physics, Ludwig-Maximilians-Universit\"at M\"unchen, 80333 M\"unchen, Germany}

\author{Guillaume Roux}
\affiliation{LPTMS, CNRS, Univ. Paris-Sud, Universit\'e Paris-Saclay, 91405 Orsay, France}

\author{I.~P.~McCulloch}
\affiliation{Centre for Engineered Quantum Systems, The University of Queensland, Brisbane, QLD 4072, Australia}

\author{Karyn Le Hur} 
\affiliation{Centre de Physique Th\'eorique, Ecole Polytechnique, CNRS, Universit\'{e} Paris-Saclay, F-91128 Palaiseau, France}
\date{\today}

\begin{abstract}
We study hard core bosons on a two leg ladder lattice under the orbital effect of a uniform magnetic field. At densities which are incommensurate with flux, the ground state is a Meissner state, or a vortex state, depending on the strength of the flux. When the density is commensurate with the flux, analytical arguments predict the possibility to stabilize a ground state of central charge $c = 1$, which is a precursor of the two-dimensional Laughlin state at $\nu=1/2$. This differs from the coupled wire construction of the Laughlin state in that there exists a nonzero backscattering term in the edge Hamiltonian.  By using a combination of bosonization and density matrix renormalization group (DMRG) calculations, we construct a phase diagram versus density and flux from local observables and central charge. We delimit the region where the finite-size ground-state displays signatures compatible with this precursor to the Laughlin state. We show how bipartite charge fluctuations allow access to the Luttinger parameter for the edge Luttinger liquid corresponding to the precursor Laughlin state. The properties studied with local observables are confirmed by the long distance behavior of correlation functions. Our findings are consistent with an exact-diagonalization calculation of the many body ground state transverse conductivity in a thin torus geometry for parameters corresponding to the  precursor Laughlin state. The model considered is simple enough such that the precursor to the Laughlin state could be realized in current ultracold atom, Josephson junction array, and quantum circuit experiments.
\end{abstract}

\maketitle

\section{Introduction}
\label{Sec:I}
\begin{figure}[t!]
  a)\includegraphics[width=0.90\linewidth]{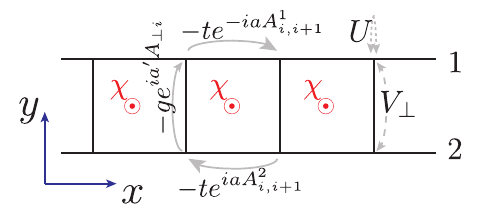}
  b)\includegraphics[width=0.90\linewidth]{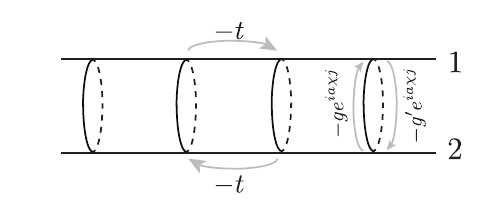}
  \caption{\label{Fig:L}(Color online) The two lattices considered in this work: a) Kinetic and interaction terms in~Eq.~(\ref{Eq:H}). The DMRG study of Sec.~\ref{Sec:N} is concerned with $g=t$ and $V_\perp = 0$ in the dilute limit of less than 1/2 particles per rung. b) Thin torus lattice corresponding to $H(0)$ in Eq.~(\ref{Eq:H_thetay}).}
\end{figure}

Quasi one--dimensional lattices in a ladder geometry with manifest time reversal symmetry breaking have been realized recently with ultracold atoms. Methods use internal atomic states to create an additional synthetic dimension, in which the orbital effect is analogous to spin--orbit coupling \cite{celi_et_al_2014,cooper_rey_2015,pagano_et_al_2014,mancini_et_al_2015, stuhl_et_al_2015}, or Raman assisted tunneling in an optical lattice \cite{aidelsburger_et_al_2013,miyake_et_al_2013,aidelsburger_et_al_2014,atala_et_al_2014,jimenez-garcia_et_al_2012,struck_et_al_2012}. These experiments probe cyclotron or skipping orbits at the edge of the sample \cite{stuhl_et_al_2015, mancini_et_al_2015}, quantum Hall transport \cite{aidelsburger_et_al_2014, miyake_et_al_2013}, as well as the Meissner state to vortex state transition\cite{orignac_giamarchi_2001} of weakly interacting Rubidium atoms\cite{atala_et_al_2014}. The latter transition is known to occur in Josephson junction ladders \cite{kardar_1986}. A typical geometry of such experiments is shown in Fig.~\ref{Fig:L}a), which represents the two--leg ladder placed in a uniform magnetic field, our focus in this work.

Originally discovered in two-dimensional electron gases \cite{tsui_et_al_1982}, the fractional quantum Hall effect has eluded implementation in quantum simulators, despite multiple theoretical proposals suitable for ultracold atoms \cite{palmer_jaksch_2003,sorensen_et_al_2005,hafezi_et_al_2007,hormozi_et_al_2012,cooper_dalibard_2013,yao_et_al_2013,Sterdyniak2014}, photonic systems \cite{hafezi_et_al_2013-2,kapit_et_al_2014}, Jaynes--Cummings--Hubbard in coupled cavity arrays \cite{cho_et_al_2008, hayward_et_al_2012, noh_angelakis_2016}, circuit quantum electrodynamics \cite{koch_et_al_2010,hartmann_2016}, or circuit QED arrays of microwave cavities \cite{anderson_et_al_2016} . Recent proposals have been put forth for cylindrical geometries \cite{lacki_et_al_2015, taddia_et_al_2016}. A recent experiment demonstrates both a synthetic magnetic field for the photons hopping in a three-qubit loop with periodically modulated couplers, and repulsive interactions mediated by the qubits\cite{roushan_et_al_2016}, which forms a scalable platform for fractional quantum Hall states of bosons.

There exist classifications of topological phases based on their coupled wire construction \cite{teo_kane_2014, neupert_et_al_2014,*meng_et_al_2015,*huang_et_al_2016}, in which bulk degrees of freedom are selectively gapped by appropriate couplings. The first example of this is the coupled wire construction of the Laughlin state \cite{kane_et_al_2002}. Interacting topological phases in quasi one dimensional geometry can be described by low--energy field theory, \textit{i.e.} bosonization \cite{giamarchi_2003}. We show in this paper that simple quantities, such as bipartite charge fluctuations\cite{song_et_al_2010,song_et_al_2011,petrescu_et_al_2014}, bridge between the field theory and experiment or numerics.

Recent investigations into fermions on the simplest geometry of two leg ladders (\textit{i.e.} systems composed of two coupled Luttinger liquids) satisfying filling fraction $\nu=1/m$ ($m=1,3,...$) reveal that the topological phase on the ladder is manifest in local observables, such as singularities in the dependence of antisymmetric chiral current on flux \cite{cornfeld_sela_2015}.  We pursue in this paper the analogous problem of a tight binding model of hard core particles with bosonic statistics at arbitrary filling and in uniform flux. It is known that the Laughlin state, which in two dimensions on a torus corresponds to a two-fold degenerate ground state that cannot be resolved by local observables, becomes a pair of charge density waves in the thin torus geometry of Fig.~\ref{Fig:L}b) \cite{seidel_et_al_2005,grusdt_hoening_2014}. Using bosonization methods we characterize the bulk and the edge excitations of the precursor to the Laughlin state on a ladder. We then identify, using numerical techniques, the region corresponding to the Laughlin phase on the phase diagram versus flux and density. This phase diagram was previously shown to contain magnetic (Mott) orders at commensurate fillings, a vortex phase at high flux, and the Meissner state at low flux \cite{piraud_et_al_2014, petrescu_le_hur_2015}, and argued to contain the Laughlin state at fillings commensurate to flux based on bosonization arguments \cite{petrescu_le_hur_2015}. Novel Mott insulating states at 1 particle per unit cell were shown to coexist with the vortex or Meissner currents \cite{petrescu_le_hur_2013, piraud_et_al_2014,petrescu_le_hur_2015}.

Our focus is an experimentally feasible low dimensional geometry in which the precursor of a topologically ordered state can be identified. This resembles the coupled wire construction of the Laughlin state by Kane \textit{et al.} \cite{kane_et_al_2002, teo_kane_2014}, with an essential difference: that our system is composed of only two wires, and consequently edge channels are not free of backscattering terms. Nonetheless, we show that signatures of the interwire couplings that generate the Laughlin state in the limit of infinitely many wires are still visible in this quasi one--dimensional geometry, embodied by a ground state of central charge $c=1$. Local observables and correlation functions allow us to distinguish this state from neighboring Meissner and vortex phases; we find that bipartite fluctuations are key in making this identification. Analytical arguments are substantiated with density matrix renormalization group (DMRG) methods \cite{white_1992,*white_1993} and exact diagonalization \cite{bauer_et_al_2011}. We have used two independent implementations of DMRG, in a numerically challenging study.

The remainder of this paper is organized as follows. Section~\ref{Sec:M} contains the model and summarizes the phase diagram obtained from analytical arguments. Section~\ref{Sec:N} contains the numerically obtained phase diagram, based on studies of central charge, local observables, and correlation functions. In Sec.~\ref{Sec:D} we describe the ground states in a thin torus geometry which generalizes the ladder geometry, and calculate the many body ground state Hall conductivity for parameters pertaining to the Laughlin state. Section~\ref{Sec:C} contains our conclusions. Further technical details are presented in a number of appendices. Appendix~\ref{App:EquivB} covers the bosonization of the ladder Hamiltonian. Appendix~\ref{App:ET} is dedicated to the details of the edge theory in the topological phase. Appendix~\ref{App:FTC} covers conventions for Fourier transforms used in the main text, and Appendix~\ref{App:CC} is dedicated to more details on the fits of central charges. Appendix~\ref{App:GSD} contains details on ground state degeneracy in the thin torus limit.

\section{Model and previous results}
\label{Sec:M}
We consider the following tight--binding model of bosons on a ladder--shaped lattice of $L$ unit cells, each composed of two sites, in a uniform magnetic field that pierces the plane of the ladder:
\begin{eqnarray}
\label{Eq:H}
H &=& -t \sum_{\alpha=1}^2 \sum_{i=1}^{L-1}  e^{i a A^\alpha_{i,i+1}}b_{\alpha, i}^\dagger b_{\alpha , i+1} + \text{H.c.} \nonumber \\
&& - g \sum_{i=1}^{L} e^{-i a' A_{\perp i}} b^\dagger_{2, i} b_{1, i} + \text{H.c.}  \nonumber \\
&&+ \frac{U}{2} \sum_{\alpha=1}^{2}\sum_{i=1}^{L} n_{\alpha, i} (n_{\alpha, i} - 1) + V_\perp \sum_{i=1}^{L} n_{1, i}n_{2, i}.
\end{eqnarray}
The geometry of the lattice and the energy scales of the model are summarized in Fig.~\ref{Fig:L}a). The summation indices in the horizontal direction correspond to open boundary conditions. The operator $b_{\alpha,i}^\dagger$ creates a particle with bosonic statistics at the $i^\textit{th}$ site on the $\alpha^\textit{th}$ chain. We consider hard core bosons with $U \to \infty$ amounting to a constraint on the local Hilbert space $n_{\alpha,i}(n_{\alpha,i} - 1)=0$. Note that Eq.~(\ref{Eq:H}) is equivalent to a spin--1/2 XXZ Hamiltonian on the ladder through the Matsubara--Matsuda\cite{matsubara_matsuda_1956,*batyev_braginskii_1984} mapping 
\begin{equation}
  \label{Eq:SpinMapping}
  b_{\alpha,i}^\dagger \to S^+_{\alpha,i},\; n_{\alpha,i} \to  S^z_{\alpha,i} + \frac{1}{2}.
\end{equation}

We denote the intrachain hopping matrix element by $t$, and the interchain hopping matrix element by $g$. The Peierls phases in Eq.~(\ref{Eq:H}) provide uniform flux $a\chi$ per plaquette, satisfying for all $i$
\begin{equation}
\label{Eq:chi}
a\chi = a A_{i,i+1}^{1} + a' A_{\perp,i+1} - a A_{i,i+1}^{2} - a' A_{\perp,i}, 
\end{equation}
which is the lattice version of the curl of the gauge field. The units of $\chi$ are inverse length, such that the number of flux quanta through the two leg ladder lattice depicted in Fig.~\ref{Fig:L}a) is $(L-1) a \chi / (2\pi)$. We also define the linear density, or the number of bosons per unit cell, in units of the inverse lattice constant 
\begin{equation}
\label{Eq:n0}
n_0 \equiv \langle \sum_{\alpha,i} n_{\alpha,i} \rangle/(La).
\end{equation}
Hereafter, model~(\ref{Eq:H}) is considered at fixed density. Our aim is the ground state phase diagram versus $(n_0,\chi)$.  

It is possible to define a filling fraction in terms of Eqs.~(\ref{Eq:chi}) and~(\ref{Eq:n0}), as commonly introduced for systems in a uniform magnetic field. This reduces to the ratio of the mean number of particles per site, and the mean number of flux quanta per square plaquette,
\begin{equation}
  \nu = \frac{n_0/2}{\chi/(2\pi)},
\end{equation}
where the numerator is the linear density per leg, $n_0/2$. In this work we will pay special attention to points on the phase diagram in the vicinity of $n_0 = \chi/(2\pi)$, \textit{i.e.} $\nu = 1/2$. In Sec.~\ref{Sec:D} we will consider closing the boundary in the $y$ direction of the lattice periodically, as shown in Fig.~\ref{Fig:L}b), while maintaining flux $a\chi$ per square plaquette. For this geometry as well the filling fraction is $\nu=1/2$.  

The phases studied below are distinguishable by local observables. In particular, the current operator is obtained from the Heisenberg equation of motion for particle number, $\dot{n_s} = i [ H, n_{s}] = \sum_{s' \neq s} j_{s',s}$, where $j_{s',s}$ denotes the number of particles per unit time transferred from site $s'$ to site $s$ ($s$ is an aggregate index for both horizontal and vertical coordinates). We rearrange the current operators at a site in linear combinations suitable for our study. The antisymmetric current operator is
\begin{eqnarray}
  \label{Eq:JParI}
  j^-_{i,i+1} &=& - i t b_{1,i}^\dagger b_{1,i+1} e^{i a A_{i,i+1}^1} + i t b_{2,i}^\dagger b_{2,i+1} e^{i a A_{i,i+1}^2} + \text{H.c.} \nonumber \\
  &\equiv& j^1_{i,i+1} - j^2_{i,i+1}.
\end{eqnarray}
The perpendicular current operator is
\begin{eqnarray}
  \label{Eq:JPerI}
  j_{\perp,i} = - i g b_{2,i}^\dagger b_{1,i} e^{-i a' A_{\perp,i}} + \text{H.c.}
\end{eqnarray}
From these we define the average antisymmetric current
\begin{equation}
  j^-_\parallel = \frac{1}{L} \left[ j_{\perp,1} - j_{\perp,L} + \sum_{i=1}^{L-1} j^{-}_{i,i+1} \right].
\end{equation}

In our analytical study of ground state orders in~\ref{Sec:B}, we consider the model~(\ref{Eq:H}) for repulsive interchain interaction $V_\perp \geq 0$. The numerical solution of Sec.~\ref{Sec:N} focuses on hard core bosons with only contact interaction, \textit{i.e.} $V_\perp = 0$, which captures the essential ground state properties for small densities.

We give an account of previous results for Eq.~(\ref{Eq:H}) at finite $U$. At $\chi=0$ and $V_\perp = 0$, and $n_0=2/a$, the ground state transitions from Mott insulator to superfluid as $g$ increases \cite{donohue_giamarchi_2001}. At arbitrary boson filling and uniform flux there is a transition from the low field Meissner phase to a high field vortex phase \cite{orignac_giamarchi_2001}, reminiscent of type-II superconductivity. The low field model with $V_\perp=0$ at $n_0=2/a$ exhibits a superfluid with Meissner currents and a Mott insulator with Meissner currents for weak enough $U$ \cite{tokuno_georges_2014}. The ground state for $\chi=\pi/a$ at integer filling is a chiral superfluid, a chiral Mott insulator or a Mott insulator \cite{dhar_et_al_2012,*dhar_et_al_2013, tokuno_georges_2014}. In the weakly interacting limit, this supports a staggered pattern of quantized orbital current vortices \cite{lim_et_al_2008,*lim_et_al_2010}.  For $\chi=0$, $V_\perp = 0$ and hard core bosons, the ground state was shown to be a rung Mott insulator at half-filling \cite{crepin_et_al_2011}. At arbitrary flux, and $n_0=1/a$, Meissner phase--Mott insulator or vortex phase--Mott insulator phases are possible due to the effective decoupling of the relative (Josephson) phase of the two legs of the ladder and the total charge \cite{petrescu_le_hur_2013}. Magnetism of the unit filled Mott phase in abelian and non-abelian gauge fields has been studied in Ref.~\onlinecite{piraud_et_al_2014-1} using DMRG. Vortex lattice states, in which the discrete lattice translation symmetry is spontaneously broken, exist for $U$ finite \cite{greschner_et_al_2015}. The two leg ladder has other such symmetry broken states: charge density waves, as well as the biased ladder state which breaks the symmetry between the two chains \cite{greschner_et_al_2016}.  

A comprehensive phase diagram versus $n_0$ and $\chi$ and interchain tunneling was obtained using the density matrix renormalization group \cite{piraud_et_al_2014}. At $2\pi n_0 = \chi$ a gapless ground state distinct from Meissner and Vortex phases was predicted \cite{petrescu_le_hur_2015}. This state matches a low--dimensional coupled wire construction of the Laughlin $\nu=1/2$ state \cite{teo_kane_2014}. Here we complement field theoretic arguments with a quantitative delimitation of this low--dimensional precursor to Laughlin phase from neighboring Meissner and vortex phases. 

\subsection{Phase diagram from bosonization}
\label{Sec:B}
In this section we summarize the phase diagram\cite{petrescu_le_hur_2015} of the continuum limit of model~(\ref{Eq:H}) obtained from the renormalization group flow equations of a bosonized \cite{giamarchi_2003} version of Eq.~(\ref{Eq:H}), in which the interchain coupling $g$ is a perturbation, \textit{i.e.} $g \ll t$. Although our numerical analysis (Sec.~\ref{Sec:N}) is performed away from this regime, namely taking $g = t$, perturbative RG allows one to understand the ordering tendency of the ground state in the thermodynamic limit.

For Eq.~(\ref{Eq:H}) with hardcore bosons $U \to \infty$, equivalently spin--$1/2$ degrees of freedom, one possibility for bosonization is to transform to fermions with the Jordan--Wigner transformation \cite{jordan_wigner_1928},
\begin{eqnarray}
  S_{\alpha,i}^{+} &\to& c_{\alpha,i}^\dagger e^{i \pi \sum_{j=1}^{n-1} c_{\alpha,j}^\dagger c_{\alpha,j}},
  \nonumber \\
  S_{\alpha,i}^z + \frac{1}{2} &\to& c_{\alpha,i}^\dagger c_{\alpha,i},
\end{eqnarray}
then derive the continuum theory of the Dirac fermions at the Fermi surface\cite{giamarchi_2003}. With the inclusion of the string operator in the expression above, it turns out that this is in fact equivalent (see App.~\ref{App:EquivB}) to taking the continuum limit directly in Eq.~(\ref{Eq:H}) by considering bosonic fields $\psi^\alpha(x) = b_{\alpha, j} / \sqrt{a} \left( = S^-_{\alpha,j}/\sqrt{a} \right)$, with $x = j a$, corresponding to each chain, expressed further as
\begin{equation}
\label{Eq:BosonOp}
(\psi^\alpha)^\dagger(x) = \sqrt{n_0^\alpha - \frac{1}{\pi} \nabla \phi^{\alpha}(x)}  \sum_p e^{ i 2 p ( n^\alpha_0 \pi - \phi^\alpha)} e^{- i \theta^\alpha(x) }, 
\end{equation} 
where $p$ runs over all integers. $\theta^\alpha(x)$ is a phase variable, while $\phi^\alpha(x')$ measures density deviations in chain $\alpha$: $\delta n^\alpha \equiv n^\alpha - n_{0}^\alpha = -\frac{1}{\pi} \nabla \phi^\alpha$. The mean densities satisfy $n_0^{1,2}=n_0/2$.  

The canonical commutation relation between the phase field and the density field
\begin{equation}
\label{Eq:A}
\left[ \phi^\alpha ( x ), \theta^\beta( x' )  \right] = i \frac{\pi}{2} \delta_{\alpha\beta} \text{sgn}( x' - x )
\end{equation} 
holds, and the following transformation of the fields is canonical
\begin{equation}
\label{Eq:Rotation}
\theta^\pm = (\theta^1 \pm \theta^2)/\sqrt{2} , \; \phi^\pm = (\phi^1 \pm \phi^2)/\sqrt{2}.
\end{equation}
This rotation is motivated by the interchain Josephson effect \cite{orignac_giamarchi_2001}, which occurs in the relative phase $\theta^-$, as discussed below. 

Equation~(\ref{Eq:H}) becomes, in the continuum limit,
\begin{equation}
\label{Eq:HWB}
\mathcal{H} = \mathcal{H}_0^+ + \mathcal{H}_0^- + \mathcal{H}_{SG}.
\end{equation}
The first two terms are Luttinger liquid (phonon) contributions
\begin{eqnarray}
\label{Eq:LLp}
\mathcal{H}_0^+ = \frac{v^+}{2\pi} \int dx \left[ K^+ (\nabla \theta^+)^2 + \frac{1}{K^+} (\nabla \phi^+)^2  \right], \\
\label{Eq:LLm}
\mathcal{H}_0^- = \frac{v^-}{2\pi} \int dx \left[ K^- (\nabla \theta^- )^2 + \frac{1}{K^-} (\nabla \phi^-)^2  \right].
\end{eqnarray}
The sound velocities and Luttinger parameters are defined by
\begin{equation}
\label{Eq:vKWB}
v^\pm = v \left[1 \pm V_\perp K a/(\pi v)\right]^{1/2}, v^\pm K^\pm = vK
\end{equation}
in terms of $v$ and $K$ of the decoupled chains (see App.~\ref{App:EquivB}). $K$ depends on the nature of intrachain interactions: $1 < K$ for repulsive interactions, $K = \infty$ for free bosons, $K<1$ for repulsive long-range interactions, and $K = 1$ for the Tonks limit considered here.  Equations~(\ref{Eq:vKWB}) hold for weak interchain coupling.

Equation~(\ref{Eq:HWB}) contains a sine-Gordon Hamiltonian arising from the interchain coupling 
\begin{eqnarray}
\label{Eq:HSG}
&&\mathcal{H}_{SG} = - 2 g \sqrt{n_0^1 n_0^2} \int dx  \cos( - \sqrt{2} \theta^-  + \chi x ) \times \\
&&\left[ 1 + 2\cos\left( 2\pi  n_0^1 x - 2 \phi^1 \right)  \right] \, \left[ 1 + 2\cos\left( 2\pi  n_0^2 x - 2 \phi^2 \right)  \right].  \nonumber
\end{eqnarray}
The flux per plaquette, $a\chi$, and the mean boson density, $n_0^\alpha$, select terms in $\mathcal{H}_{\textit{SG}}$ whose renormalization group flows \cite{giamarchi_2003} are towards strong coupling. Such relevant contributions control the ground state in the thermodynamic limit.

We make a brief note on local observables. Current density operators are obtained from the Heisenberg equation for the density operator $n^{1}(x)-n^{2}(x) = \text{const} - \frac{1}{\pi} \sqrt{2} \nabla \phi^{-}$:
\begin{equation}
  \frac{d}{dt}\left( n^{1}(x)-n^{2}(x) \right) = i [ \mathcal{H}, n^{1}(x)-n^{2}(x) ],
\end{equation}
from which currents $j_\perp(x)$ and $j^-_\parallel(x)$ are obtained using the same conventions as those introduced in Eqs.~(\ref{Eq:JParI}) and~(\ref{Eq:JPerI}) for lattice currents. Throughout this work we will refer to current density operators as ``currents''.

We focus on densities $a n_0 \leq 0.5$, such that terms sustaining charge density waves, such as $\cos( 2 \pi n_0 x - 2 \sqrt{2} \phi^+ )$, are irrelevant. In particular, density $n_0=1/a$ stabilizes the rung Mott insulator phase\cite{petrescu_le_hur_2015}. Away from such commensurate densities, Eq.~(\ref{Eq:HSG}) predicts three phases: the Meissner phase, the vortex phase, and the Laughlin phase. 

At small $\chi$, and eliminating the possibility of commensuration between $\chi$ and $2\pi n_0$,
\begin{equation}
  \label{Eq:HSGM}
  \mathcal{H}_{SG} = - g n_0 \int dx  \cos( - \sqrt{2} \theta^-  + \chi x ).
\end{equation}
This potential pins the gauge invariant bosonic phase $-\sqrt{2}\theta^- + \chi x$, inducing a gap in the ``-'' sector. This gap has power law dependence on the interchain coupling
\begin{equation}
\label{Eq:Deltam}
\Delta^- \sim \frac{v}{a} \left( \frac{ga}{v} \right)^{\frac{1}{2-\frac{1}{2 K^-}}},
\end{equation}
which implies exponentially decaying rung current two point correlation functions, on a distance that scales with the inverse gap, $\xi^- \propto 1/\Delta^-$, 
\begin{equation}
  \label{Eq:RR}
  \langle j_\perp(x) j_\perp(0) \rangle \sim \exp(-|x|/\xi^-) \to 0 \text{ as } x/\xi^- \to \infty.
\end{equation}
The rung current density operator
\begin{equation}
  \label{Eq:JR}
  j_\perp(x) = \frac{2g}{\pi a} \sin( \sqrt{2}\theta^- - \chi x)
\end{equation} 
has vanishing ground state expectation value. The antisymmetric parallel current density, 
\begin{equation}
j_\parallel^-(x) = -vK \sqrt{2} \nabla \theta^-,
\end{equation} 
has ground state expectation value $\langle j_\parallel^-(x) \rangle = - vK \chi$, which is reminiscent of the Meissner effect \cite{orignac_giamarchi_2001}. As the ``$+$'' fields are gapless and decoupled from the dynamics of the ``$-$'' fields, this corresponds to central charge\cite{gogolin_et_al_1998} $c=1$.

For $\chi > \chi_c \sim \pi \sqrt{2} \Delta^- / (vK)$, the sine--Gordon Hamiltonian~(\ref{Eq:HSGM}) becomes irrelevant, which corresponds to a commensurate incommensurate transition \cite{pokrovsky_talapov_1980,*schulz_1980,*giamarchi_schulz_1988}. At high flux and in the absence of commensuration effects between the flux and the density, the ground state is gapless and has central charge $c=2$. This is the vortex state \cite{orignac_giamarchi_2001}. 

The situation is qualitatively distinct if there exists commensuration between flux and density. For $\chi > \chi_c$ and whenever the commensuration condition $a\left[ 2\pi n_0 \pm \chi \right] = 0 \text{ mod }2\pi$ holds, the number of particles per flux quantum is $\nu = 1/2$. For the lower sign in the commensuration condition above,
\begin{equation}
  \label{Eq:LaughlinCouplingm}
  \mathcal{H}_{\textit{SG}} = - g n_0 \int dx \cos( -\sqrt{2} \theta^- + m \sqrt{2} \phi^+ ),\;\; m = 2.
\end{equation}
This term is relevant provided that its scaling dimension satisfies the condition
\begin{equation}
  \label{Eq:SDL}
  2 > 1/(2K^{-}) + 2 K^{+}.
\end{equation}
The energetics in Eq.~(\ref{Eq:H}) can be tuned in order to satisfy Eq.~(\ref{Eq:SDL}). Starting from the perturbative limit $g \ll t$, we may obtain (App.~\ref{App:EquivB}) an estimate for the strength of repulsive interactions necessary to satisfy this inequality in the Tonks limit $U\to \infty$: $V_\perp \gtrsim 3t$. This estimate, however, is not rigorous. The Luttinger parameters $K^{\pm}$ are expected to vary with filling \cite{crepin_et_al_2011} and flux, and ultimately one needs to verify Eq.~(\ref{Eq:SDL}) from numerics, e.g. by fitting appropriate correlation functions to extract Luttinger parameters. For the dilute limit considered here, $a n_0 \leq 0.5$, rung repulsion $V_\perp$ is expected to play a less significant role, and hence only weak dependence of $K^{\pm}$ on $V_\perp$ is expected. To summarize, in the dilute limit, the aforementioned bound $V_\perp \gtrsim 3t$ appears to be too strict. Guided by these observations, and backed by numerical checks, in Sec.~\ref{Sec:N} we perform studies at $V_\perp=0$.

Provided that Eq.~(\ref{Eq:SDL}) is satisfied, the pinning in Eq.~(\ref{Eq:LaughlinCouplingm}) favors a $c=1$ gapless ground state corresponding to a gapped bulk field, represented by the linear combination $-\sqrt{2} \theta^- +  \sqrt{8} \phi^+$, decoupled from a gapless edge Luttinger liquid\cite{teo_kane_2014}. Perturbative RG in $g \ll t$ shows that the ``bulk'' gap is a power law in the interchain coupling, 
\begin{equation}
\label{Eq:LaughlinGap}
\Delta \sim \frac{v}{a} \left( \frac{g a}{v} \right)^{\frac{1}{2 - 1/(2K^-) - 2 K^+}},
\end{equation}
This gap is significantly reduced as compared to $\Delta^-$. Equation~(\ref{Eq:LaughlinGap}) gives the size of the gap above the central charge $c=1$ ground state if Eq.~(\ref{Eq:SDL}) is satisfied. If, however, Eq.~(\ref{Eq:SDL}) is not satisfied, for finite system sizes, the irrelevant operator Eq.~(\ref{Eq:LaughlinCouplingm}) may still influence the ground state properties, although in the thermodynamic limit there is no effect of the pinning potential~(\ref{Eq:LaughlinCouplingm}). We identify the $\nu=1/2$ pinning potential described in the last part of this subsection with the one responsible for opening the bulk gap in the coupled wire construction of the Laughlin state \cite{kane_et_al_2002}.

\subsection{Edge theory in Laughlin phase}
The edge Luttinger liquid corresponding to the $c=1$ Laughlin phase can be obtained by integrating the gapped field in Eq.~(\ref{Eq:LaughlinCouplingm}) (for a detailed calculation, see App.~\ref{App:ET}). The gapped bulk degree of freedom can be recast in terms of the fields
\begin{equation}
  \phi = -\frac{\theta^-}{m\sqrt{2}} + \frac{\phi^+}{\sqrt{2}}, \;  \theta =  \frac{\theta^+}{\sqrt{2}} - m\frac{\phi^-}{\sqrt{2}}.
\end{equation}
The gapless degrees of freedom are expressed in terms of the fields
\begin{eqnarray}
  \phi' =  \frac{\theta^-}{m\sqrt{2}} + \frac{\phi^+}{\sqrt{2}}, \;
  \theta' =  \frac{\theta^+}{\sqrt{2}} + m\frac{\phi^-}{\sqrt{2}}.
\end{eqnarray}
The edge excitations are described by the Luttinger liquid Hamiltonian
\begin{equation}
  \label{Eq:He}
  \mathcal{H}^e = \frac{v^e}{2\pi} \int_0^L dx \left[ K^e(\nabla \theta')^2 + \frac{1}{K^e} (\nabla \phi')^2 \right],
\end{equation}
where the edge Luttinger parameter and sound velocity are specified by (see App.~\ref{App:ET})
\begin{eqnarray}
  v^e K^e &=& \frac{vK + h^- \frac{v}{m^2 K}}{2} 
  - \frac{\left( v K - h^- \frac{v}{m^2 K} \right)^2}{2\left( v K + h^- \frac{v}{m^2 K} \right)},
\end{eqnarray}
and
\begin{equation}
  \frac{v^e}{K^e} = v \frac{ m^2 K + h^+ \frac{1}{K}}{2},
\end{equation}
with $m=2$ and $h^{\pm} \equiv 1 \pm \frac{V_\perp K a}{\pi v}$ differing from unity by an amount that measures the strength of rung repulsion. At $V_\perp=0$ we find $K^e = 2/5$ and $v^e=v$. 

The edge Luttinger liquid Eq.~(\ref{Eq:He}) contains backscattering terms, which distinguishes it from the chiral Luttinger liquid describing the edge excitations of a quantum Hall droplet \cite{wen_1995,*wen_1992}. That is, we find that
\begin{eqnarray}
\label{Eq:Hec}
\mathcal{H}^e = \frac{v}{8\pi} \int_0^L dx \big[ A_{RR} (\nabla R)^2 + A_{LL} (\nabla L)^2 + \nonumber \\ A_{LR}(\nabla R)(\nabla L)  \big].
\end{eqnarray}
The chiral fields are defined as 
\begin{equation}
  R(x) = \phi' + \theta',\; L(x) = \theta'-\phi',
\end{equation}
and obey the algebra
\begin{equation}
  [R(x),R(x')] = i \frac{\pi}{m} \text{sgn}(x'-x) = - [L(x),L(x')].
\end{equation}
The coefficients in Eq.~(\ref{Eq:Hec}) are
\begin{eqnarray}
  A_{RR} &=& A_{LL} =  m^2 K^e + \frac{1}{K^e}, \nonumber \\
  A_{LR} &=& \frac{2}{K^e} - 2 m^2 K^e.
\end{eqnarray}
Nonzero $A_{LR}$ signifies that there is backscattering between the edge chiral fields, resulting from the traced bulk degrees of freedom.

Note that long range repulsion within each chain such that $K=1/m$ results in a chiral Luttinger liquid Hamiltonian 
\begin{equation}
  \mathcal{H}^e = \frac{mv}{4\pi} \int_0^L dx \left[ (\nabla R)^2 + (\nabla L)^2 \right].
\end{equation}
This has the same form as the chiral Luttinger liquid edge theory of a $\nu=1/m$ bulk discussed by Wen \cite{wen_1995,*wen_1992}. In the two--chain problem, long range repulsive interactions are required to cancel edge backscattering contributions occuring through the bulk. This is unlike the case of a coupled wire construction with many wires, in which backscattering amplitudes are exponentially suppressed through the gapped bulk \cite{neupert_et_al_2014}.

We conclude that interactions which render~(\ref{Eq:LaughlinCouplingm}) relevant generally induce backscattering at the edge. This is in contrast to the integer quantum Hall effect of fermions ($\nu=1$), corresponding to Eq.~(\ref{Eq:LaughlinCouplingm}) with $m=1$, which is relevant for $V_\perp=0$, and where the edge theory is a chiral Luttinger liquid with $K^e=K=1$.

The resulting edge Luttinger liquid is characterized by charge fractionalization \cite{safi_schulz_1995,*pham_et_al_2000,*le_hur_2006,*le_hur_et_al_2008,*SteinbergEtAl2008,*berg_et_al_2009}. Without edge backscattering terms, edge current noise at a quantum point contact probes the fractional charge of bulk quasiparticles \cite{kane_fisher_1994}. This has been demonstrated for a 2D electron gas \cite{de_picciotto_et_al_1997,*radu_et_al_2008} and is feasible with ultracold atom quantum point contacts \cite{brantut_et_al_2012}. In Sec.~\ref{Subsec:F} we show that the Luttinger parameter $K^e$ defining the edge Hamiltonian~(\ref{Eq:He}) can be extracted robustly from bipartite charge fluctuations obtained from numerics.

\section{Phase diagram from DMRG}
\label{Sec:N}
This section contains the phase diagram for the ground states of~(\ref{Eq:H}) obtained using the density matrix renormalization group algorithm \cite{white_1992,*white_1993}. We make the following parameter choices
\begin{equation}
  g = t = 1, \; V_\perp = 0, \; U = \infty.
\end{equation}
We delimit previously studied Meissner and vortex liquid phases\cite{piraud_et_al_2014} after which we focus our attention to a small region of the phase diagram that we argue has features which are consistent with the low--dimensional Laughlin $\nu=1/2$ ground state. 

We have used two independent implementations of DMRG in order to tackle the numerically challenging regions of the phase diagram.  We have focused on open boundary conditions in the $x$ direction. Although this choice subjects us to the effect of the boundary in the form of Friedel oscillations, it has the advantage of better convergence. We have considered system sizes up to $L=225$ and up to $1200$ states.  For the coarse phase diagram in the $(n_0,\chi)$ plane, shown in Figures~\ref{Fig:P} and~\ref{Fig:CCNV}, we have considered $L=65$ rungs, and retained up to $800$ states. Convergence was especially problematic near half a flux quantum per plaquette, $a\chi = \pi$, where $H$ in~(\ref{Eq:H}) is time-reversal invariant. We have omitted these flux values from the phase diagram, as they are beyond the scope of this study.

The section is organized as follows. Section~\ref{Subsec:P} contains the numerical phase diagram obtained from central charge and vortex number. In Sec.~\ref{Subsec:CR} we argue that the Laughlin phase is characterized by local extrema in local current operators. Bipartite fluctuations (Sec.~\ref{Subsec:F}) allow us to elucidate between two operators which compete to order the ground state. Section~\ref{Subsec:C} is dedicated to the study of correlation functions of current operators and of the vertex operator corresponding to the Laughlin state in Eq.~(\ref{Eq:LaughlinCouplingm}).

\subsection{Phases from central charge and vortex number}
\label{Subsec:P}
\begin{figure}[t!]
  \includegraphics[width=\linewidth]{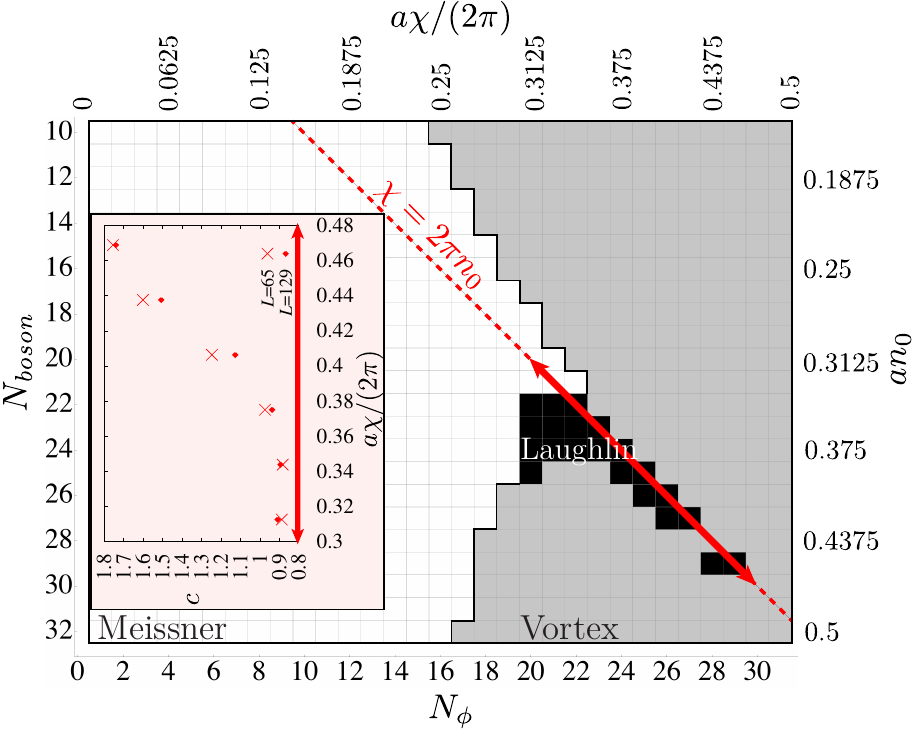}
  \caption{\label{Fig:P}(Color online) DMRG phase diagram for open boundary $L=65$ ladder versus number of particles and number of flux quanta threading the ladder (definitions in Table~\ref{Tab:P}). The dashed line corresponds to $\chi = 2\pi n_0$. Inset shows fit for central charge according to~(\ref{Eq:SF}), for points at $\chi = 2\pi n_0$ under the solid arrow: $L = 65$ (crosses) and $L=129$ (dots).}
\end{figure}

\begin{figure}[t!]
  a)\includegraphics[width=0.95\linewidth]{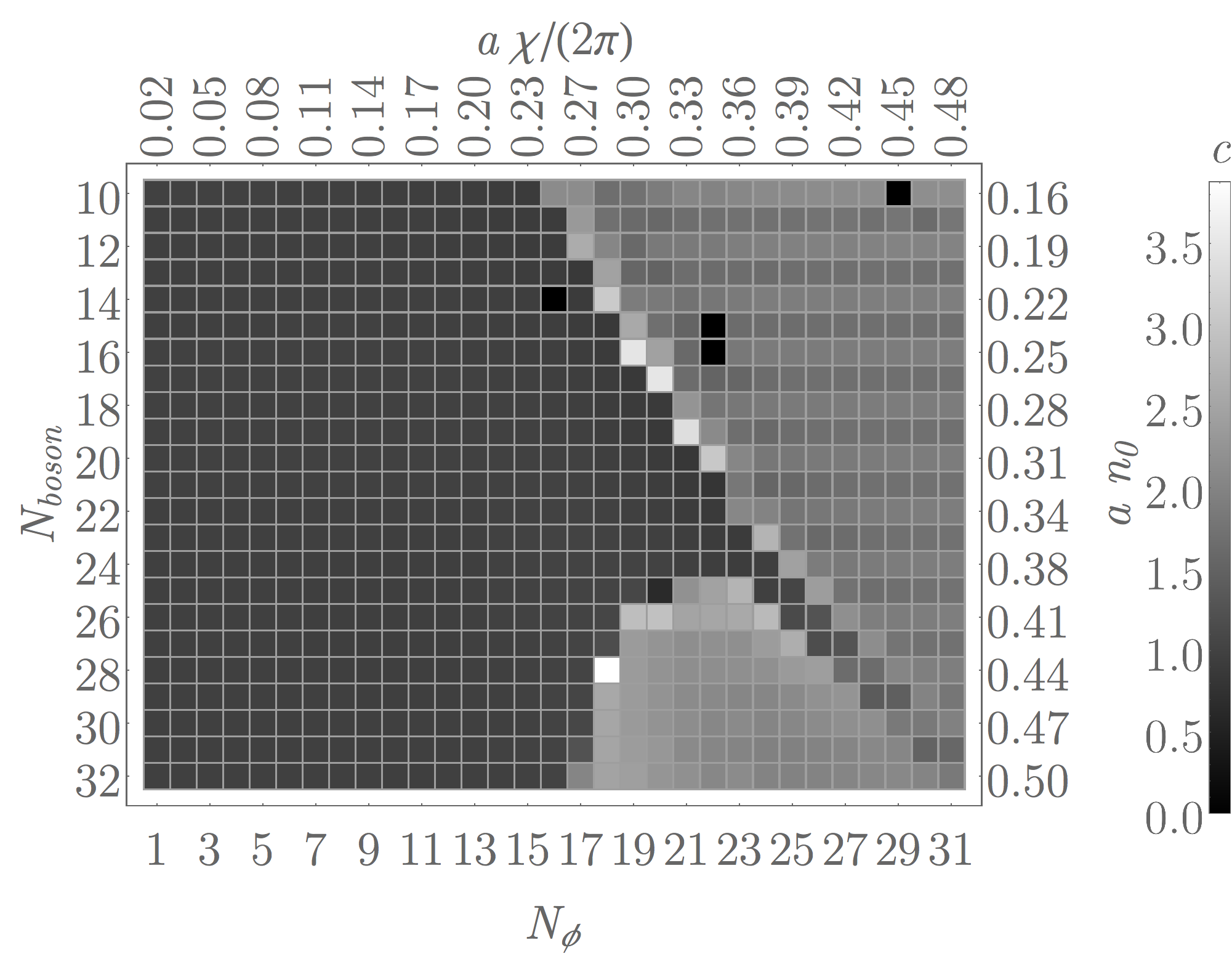}
  b)\includegraphics[width=0.95\linewidth]{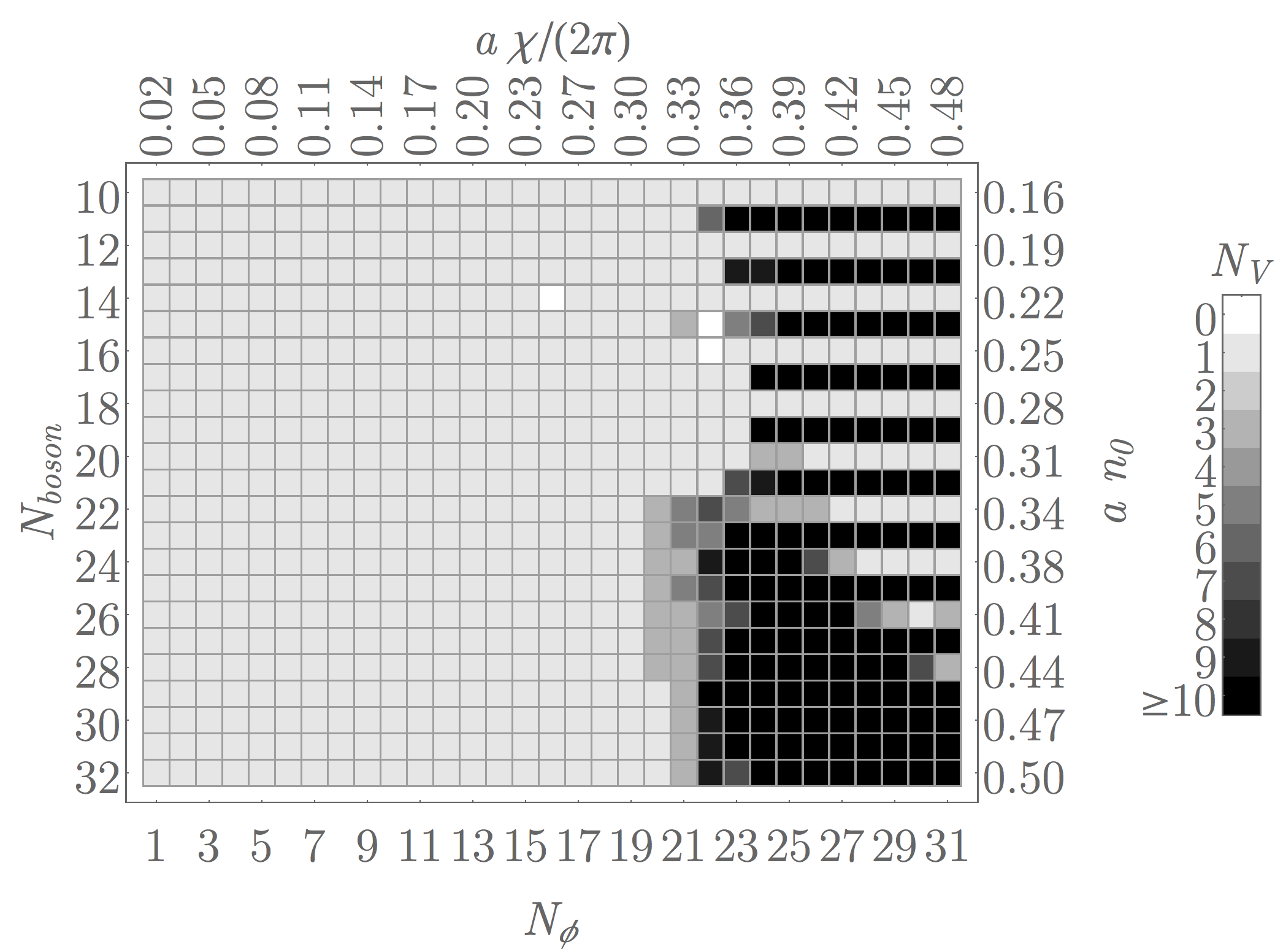}
  \caption{\label{Fig:CCNV}a) Central charge $c$ from fits of the bipartite entanglement entropy in $(n_0,\chi)$ plane obtained from~(\ref{Eq:SF}) on $L = 65$ rung ladder with OBC; b)~Vortex number $N_V$ defined in Eq.~(\ref{Eq:NV}) in the same coordinates.}
\end{figure}

We construct the phase diagram of Fig.~\ref{Fig:P} based on central charge and current expectation values, the latter through a single observable, the vortex number. These quantities are plotted in Fig.~\ref{Fig:CCNV}. The Meissner, Vortex, and Laughlin phases, together with their identification in terms of central charge and vortex number, are summarized in Table~\ref{Tab:P}. 

The central charge is obtained from the prefactor of the logarithmic contribution \cite{holzhey_et_al_1994,vidal_et_al_2003,calabrese_cardy_2009} to the bipartite entanglement entropy \cite{von_neumann_1955}. The entanglement entropy contains additional subleading oscillatory corrections
\begin{equation}
  \label{Eq:SF}
  S(j) = \frac{c}{6} \log\left[ d(j | L ) \right] + A \langle b_{1,j}^\dagger b_{1,j+1} \rangle + B,
\end{equation}
where $\log$ is the natural logarithm, and $A$ and $B$ are nonuniversal coefficients. Equation~(\ref{Eq:SF}) holds for open (Dirichlet) boundary conditions for the wavefunction, and $d( x | L) = (L/\pi) \sin( \pi x / L )$ is a compactified distance function\cite{cazalilla_2004}. (In periodic boundary conditions there is an enhancement of the logarithm prefactor $c/6 \to c/3$.)  The central charge obtained from fits using Eq.~(\ref{Eq:SF}) in a $L=65$ rung ladder is plotted in Fig.~\ref{Fig:CCNV}a). Further fits for the central charge, using finite size scaling, are detailed in App.~\ref{App:CC}. 

\begin{table}[t!]
  \begin{tabular}{|l | l  l l|}
    \hline
    Ground state       & Meissner  &  Vortex &  Laughlin \\
    \hline
    $c$                & $1$       &  $2$      &   $1$ \\
    \hline    
    $N_V$              & $1$       &  $\geq 1$ &   $>1$ \\
    \hline
  \end{tabular}
  \caption{\label{Tab:P}Criteria for the phase diagram of Fig.~\ref{Fig:P}, based on central charge and vortex number. }
\end{table}

\begin{figure}[t!]
  \includegraphics[width=\linewidth]{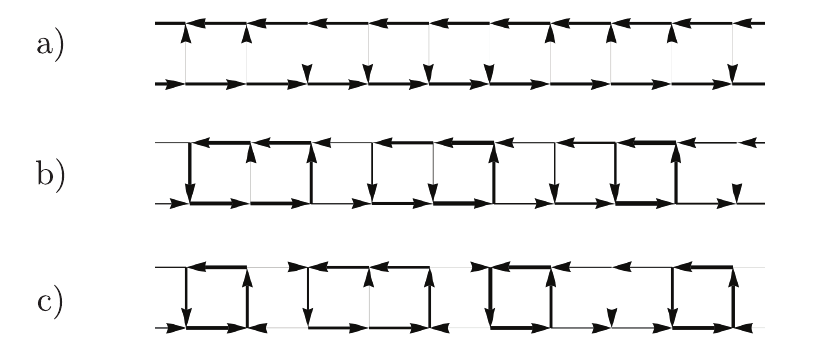}
  \caption{\label{Fig:CEV}Bond current expectation value where $a n_0=a\chi/(2\pi)$ takes the following values: a) $0.155$, b) $0.315$, c) $0.375$.}
\end{figure}

For the purpose of generating a phase diagram, we characterize the current pattern by a single number, the vortex number order parameter
\begin{equation}
\label{Eq:NV}
N_V \equiv \sum_{i=2}^{L-1} \Theta\left( - \text{sign} \frac{\langle j^1_{i,i+1} \rangle}{\langle j^1_{i-1,i} \rangle} \right) + 1,
\end{equation}
where $\Theta$ is the Heaviside step function \cite{gradshteyn_ryzhik_2007} and angular brackets denote the ground state expectation value. Note that $N_V=1$ in the Meissner state, where $\langle j_{\perp,i} \rangle=0$ for $1<i<L$ precludes a sign change in $\langle j^{1,2}_{i,i+1} \rangle$ for $1\leq i <L$ by the continuity equation. Typical expectation values of the bond current operators are shown in Fig.~\ref{Fig:CEV} for a collection of points on the Laughlin line $n_0 = \chi / (2\pi)$.

Note that the existence of sign changes in the horizontal currents $\langle j^{1}_{i,i+1} \rangle$ is indicative of the fact that the $0$ momentum (Meissner) component of the antisymmetric current becomes subdominant compared to a finite momentum, staggered, component. For example, Figure~\ref{Fig:CEV} shows local current patterns and illustrates that at high enough flux the uniform current pattern reminiscent of Meissner phase is no longer visible. The condition $N_V > 1$ is stricter than necessary to detect the commensurate--incommensurate transition from the Meissner phase to the vortex phase\cite{piraud_et_al_2014}. The transition to the vortex state is identifiable via the discontinuity in central charge (see Fig.~\ref{Fig:CCNV}a)), and via the development of a nonzero momentum peak in the antisymmetric current~\cite{piraud_et_al_2014}. Numerically we find that $N_V > 1$ only in a subregion of the vortex phase (which has $c=2$). The points with $N_V = 1$ inside the vortex phase are possibly a consequence of finite size: this only indicates that the Meissner current (zero momentum component of the antisymmetric current) is dominant.  Importantly, phases with $N_V > 1$ and $c=1$, as found here on the line $\chi=2\pi n_0$ for large enough $\chi$, are neither in the Meissner phase, nor vortex phase; we identify this region as the Laughlin phase.

The phase diagram of Fig.~\ref{Fig:P} is obtained based on the values of $c$ and $N_V$ (see Table~\ref{Tab:P}). The transitions from the Meissner phase to the Vortex phase and from the Laughlin phase to the Vortex phase are inferred from the central charge and the Meissner to Laughlin transition from $N_V$, as explained above. In particular, the $c=1$ phase with $N_V > 1$ at intermediate flux values $0.26 \lesssim a\chi/(2\pi) \lesssim 0.45$ in the vicinity of the line $\chi = 2\pi n_0$ is identified as the Laughlin phase.

\subsection{Antisymmetric chiral current}
\label{Subsec:CR}
\begin{figure}[t!]
  \includegraphics[trim={0.5cm 1cm 2cm 6cm},clip,width=\linewidth]{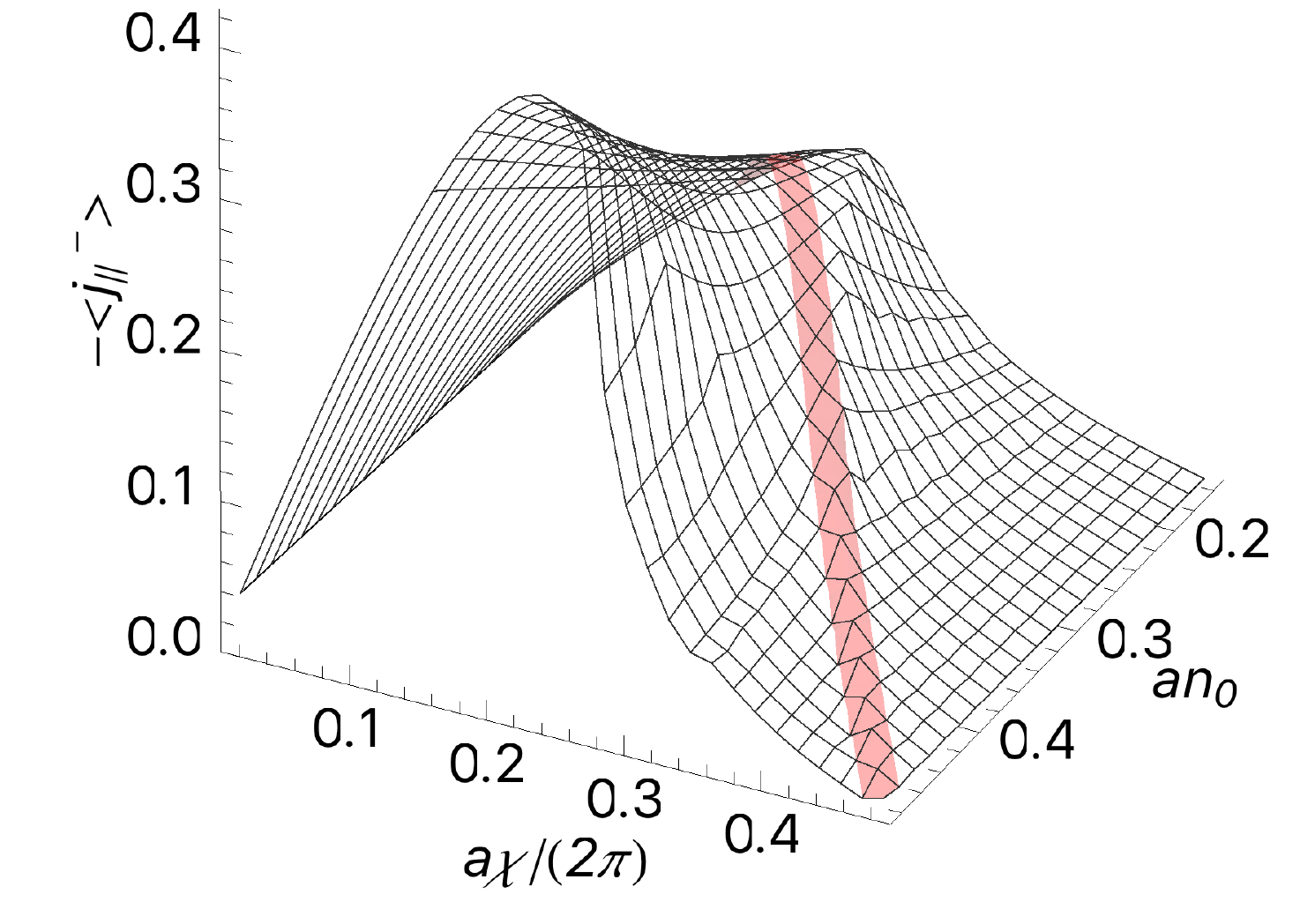}
  \caption{\label{Fig:JP} (Color online) Antisymmetric parallel current $\langle j_\parallel^- \rangle$: Meissner effect for low $\chi$.  On the dark red shaded line, corresponding to $2\pi n_0 = \chi$, there are local extrema of the current operator expectation value at large enough $\chi /(2\pi) > 0.3$. The positions of these extrema belong to the Laughlin phase as indicated by the thick red line on Fig.~\ref{Fig:P}.}
\end{figure}

At low flux values, in the Meissner phase, we expect that the antisymmetric parallel current screens the flux, $\langle j^{-}_{\parallel} \rangle \propto \sin(a \chi)$. Close to $a\chi = \pi$, the average antisymmetric parallel current is expected to vanish, as an orbital antiferromagnet pattern of staggered currents forms \cite{dhar_et_al_2012,*dhar_et_al_2013}. At $\chi = 2\pi n_0$ for flux values which are intermediate between the Meissner phase and the orbital antiferromagnet regime, we notice the formation of local extrema (see Fig.~\ref{Fig:JP}), where numerical results suggest that $\partial \langle j_\parallel^-\rangle / \partial \chi$ and $\partial \langle j_\parallel^-\rangle / \partial n_0$ have a finite discontinuity. The local extrema of $\langle j^-_\parallel \rangle$ occur on the thick red line identified as the Laughlin phase on Fig.~\ref{Fig:P}.

Returning to our treatment in Sec.~\ref{Sec:B}, note that the pinning potential~(\ref{Eq:LaughlinCouplingm}) implies that $j_\perp(x) = \frac{2g}{\pi a} \sin( \sqrt{2}\theta^- - 2\sqrt{2} \phi^+ )$ has vanishing ground state expectation value; moreover, that 
\begin{equation}
  \label{Eq:LLC}
\langle j_\parallel^- \rangle = -vK \sqrt{2}\langle \nabla \theta^- \rangle = - 2 \sqrt{2} v K \langle \nabla \phi^+ \rangle,
\end{equation}
\textit{i.e.} at any position $\langle j_\parallel^-(x) \rangle$ is proportional to total density fluctuations. Then necessarily the antisymmetric parallel current integrated over the length of the system vanishes, contrary to what is found numerically in Fig.~\ref{Fig:JP}. We attribute this discrepancy to the neglect of Fermi sea contributions in the bosonization treatment, in agreement with a recent analytical argument for the $\nu=1$ integer quantum Hall state in the same geometry \cite{cornfeld_sela_2015}.

\begin{figure}
  \includegraphics[width=0.49\linewidth]{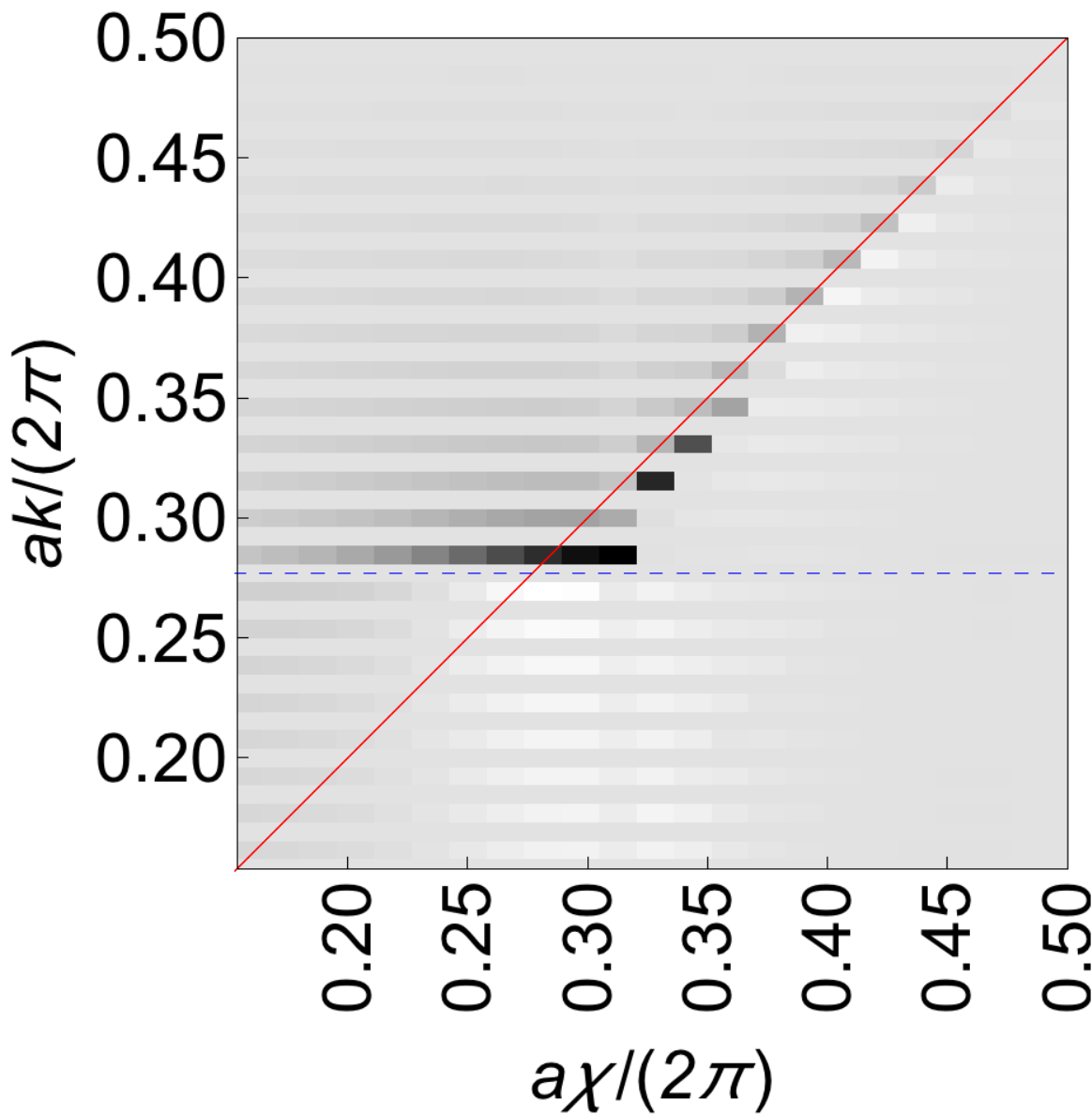}
  \includegraphics[width=0.49\linewidth]{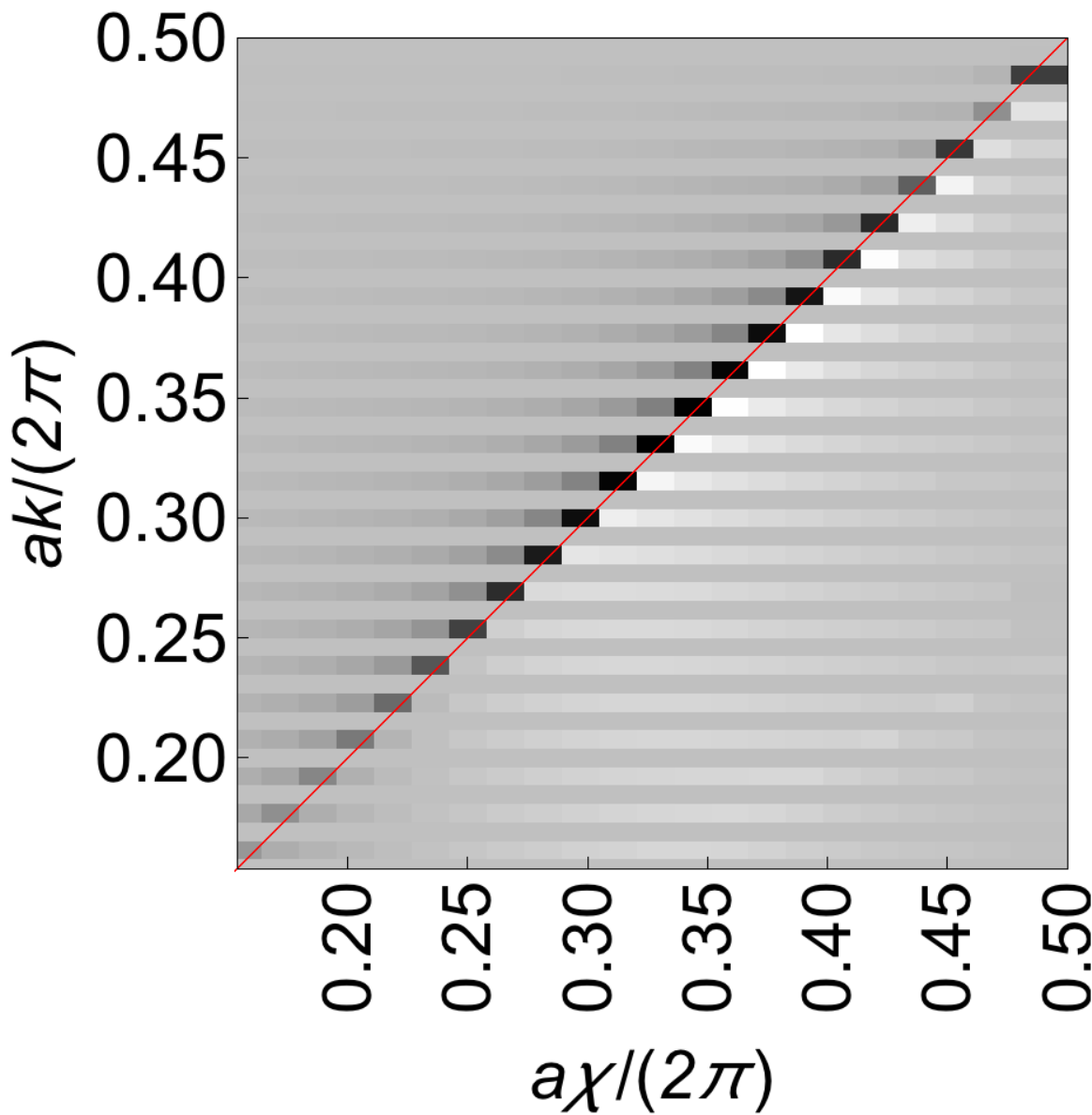}    
  \caption{\label{Fig:FT}(Color online) $\langle \delta j_\parallel^-(k)\rangle$ as a function of flux $\chi$ in arbitrary units (darker is more negative). Left: At fixed density (e.g. here $a n_0\approx 0.28$), the extremum is at $k=2\pi n_0$ for $\chi < 2\pi n_0$ and at $k=\chi$ for $\chi >2\pi n_0$. Dashed dark (blue) line corresponds to $k/(2\pi)=n_0$, while continuous dark (red) line corresponds to $k=\chi$. Right: When $\chi/(2\pi) = n_0$, there is a single nonzero minimum at $k/(2\pi)=\chi/(2\pi)$.}
\end{figure}

A complementary signature of the phase transitions are peaks in the Fourier transforms of the current operator. For open boundary conditions, we define the Fourier transform
\begin{equation}
 f[k] = \sqrt{\frac{2}{L+1}} \sum_{j=1}^{L} \sin\left(j k \right) f_j,
\end{equation}
where $k = n_k \pi / (L+1)$ with $n_k=1,...,L$. Note that the basis functions appropriately vanish at $j=0$ and $j=L+1$. We note the following for $\langle j_\parallel^- \rangle[k]$: In the Meissner phase, it has a peak in the vicinity of $k=0$, as well as a smaller nonzero momentum peak at $k\approx 2\pi n_0$. In Fig.~\ref{Fig:FT} we consider $\delta \langle j_\parallel^- \rangle[k]$ in which the mean value ($k=0$ component) is subtracted. In the vortex phase, $\delta \langle j_\parallel^- \rangle [k]$ peaks near $k = \chi$. In summary, apart from the Meissner peak at zero momentum, the position of the peak in the Fourier transform of the antisymmetric chiral current is subject to the competition between ordering at $2\pi n_0$ (deep Meissner phase) or at $\chi$ (deep vortex phase). At values of flux close to $\chi = 2\pi n_0$ in the vortex phase, for example in the $c=2$ phase around $n_0 (L-1) = 27$ on Fig.~\ref{Fig:P}, the peak may be situated at an intermediate value between $\chi$ and $2\pi n_0$. Consequently, whenever the flux per plaquette equals the density $\chi=2\pi n_0$ the nonzero momentum peak is at their common value (see Fig.~\ref{Fig:FT}).

It is noteworthy that the Fourier transform of the deviation of particle number from its mean, $\langle n_{\alpha,i} \rangle - n_0/2$, seems to be within a numerical factor of $\langle j^-_{i,i+1}\rangle$ even away from $n_0 = \chi/(2\pi)$. This shows that the criterion~(\ref{Eq:LLC}) cannot delimit the Laughlin phase. We infer that this pinning of density to antisymmetric currents holds even away from $\chi = 2\pi n_0$ because, for finite system size, the irrelevant operator in Eq.~(\ref{Eq:LaughlinCouplingm}) may still control ordering in the ground state.

\subsection{Bipartite fluctuations}
\label{Subsec:F}
Bipartite fluctuations of particle number contain information about quantum entanglement \cite{song_et_al_2010,song_et_al_2011,petrescu_et_al_2014}. In the gapless phases studied here, bipartite particle number fluctuations can be used to extract Luttinger parameters. Bipartite fluctuations may be thought as originating from the logarithm of a correlation function of exponentials of fields: for a Gaussian action and an arbitrary field $\eta$, exponential correlation functions are determined by the two point correlation function only \cite{giamarchi_2003}
\begin{equation}
  \left\langle  e^{i \eta (x)} e^{- i \eta(y)} \right \rangle = e^{-\frac{1}{2} \langle [\eta(x) - \eta(y) ]^2 \rangle}.
\end{equation}
In this section $\eta$ will be one of the density fields $\phi^{\pm}$. Taking a logarithm of the above relates a correlator which is relatively hard to obtain in numerics (the left hand side), to a simpler quantity on the right hand side. Fluctuations are readily available in the DMRG routine via the evaluation of local boson densities, unlike correlation functions of exponentials of fields such as $\phi^\pm(x)$ (string operator correlation functions), which introduce computational complications. The latter will be discussed in the next Subsec.~\ref{Subsec:C}. In this section, we derive analytical forms implied by the field theory of Sec.~\ref{Sec:B}, then proceed to numerical fits.

Bipartite particle number fluctuations are measured by the connected correlation function
\begin{equation}
  \mathcal{F}^\pm(\ell) \equiv \langle \left[ N^\pm(\ell) \right]^2 \rangle_\text{conn}.
\end{equation}
We have introduced the total and relative particle numbers in a subsystem comprised of the first $\ell$ rungs:
\begin{equation}
  N^\pm(\ell) = \sum_{i=1}^{\ell} \left( n_{1i} - n_{2i} \right).
\end{equation}
With our definitions in Sec.~\ref{Sec:B},
\begin{equation}
  N^\pm( \ell ) = - \frac{\sqrt{2}}{\pi} \left[ \phi^\pm(\ell) - \phi^\pm(0)  \right],
\end{equation}
leading to 
\begin{eqnarray}
  \label{Eq:Fpm}
  \mathcal{F}^\pm(\ell) &=& \frac{2}{\pi^2} \langle \left[ \phi^\pm(\ell) - \phi^\pm(0) \right]^2 \rangle_\text{conn} \nonumber \\
  &=& \frac{2}{\pi^2} \langle \phi^\pm(\ell) \phi^\pm(\ell) \rangle_\text{conn} + \frac{2}{\pi^2} \langle \phi^\pm(0) \phi^\pm(0) \rangle_\text{conn}  \nonumber \\ &&-  \frac{4}{\pi^2} \langle \phi^\pm(\ell) \phi^\pm(0) \rangle_\text{conn}.
\end{eqnarray}
The correlation functions of the fields $\phi^\pm$ in open/periodic boundary conditions are known \cite{cazalilla_2004}. 

Let us begin with the correlation functions in the vortex phase. Density two point correlation functions decay as a power law with separation
\begin{eqnarray}
  &&\langle  \nabla \phi^\pm(x) \nabla\phi^\pm(x') \rangle = \\
  &&\;\;\;\;\;- \frac{K^\pm}{2} \left[  \frac{1}{d(x-x'|2L)^2} + \frac{1}{d(x+x'|2L)^2}   \right], \nonumber 
\end{eqnarray}
or, after the indefinite integral,
\begin{eqnarray}
  \label{Eq:PhiPhiOBC}
  &&\langle  \phi^\pm(x) \phi^\pm(x') \rangle =  \\
  &&\;\;\;\;\;-\frac{K^\pm}{2} \log\left[ d(x-x'|2L)\right] + \frac{K^\pm}{2} \log\left[ d(x+x'|2L) \right]. \nonumber 
\end{eqnarray}
Regularizing $d(x|L)=\text{const}$ for $x<a$, we use the above expression~(\ref{Eq:PhiPhiOBC}) into the formula for the bipartite fluctuations, Eq.~(\ref{Eq:Fpm}), and obtain:
\begin{eqnarray}
  \label{Eq:FpmellOBC}
  \mathcal{F}_\textit{vortex}^\pm(\ell) = \frac{K^\pm}{\pi^2} \log\left[ d(\ell|L) \right] + c + d b_E(\ell),
\end{eqnarray}
where we have included subleading constant and bond energy contributions, with $b_E(\ell) = \langle b_{1,\ell}^\dagger b_{1,\ell+1} \rangle$. The logarithmic dependence can be understood as follows. The hopping term of Eq.(\ref{Eq:HSGM}) becomes irrelevant \cite{petrescu_le_hur_2015}. There is no tunneling between the chains in the thermodynamic limit, and one expects gaussian fluctuations in both total and relative charges. For finite sized systems, the Josephson term may give a linear contribution $\propto \ell$, a direct consequence of the finite size gap in the $\theta^-$ sector.

In the Meissner phase the $\theta^-$ field is gapped by the Josephson term~(\ref{Eq:HSGM}). One important point is that the particle number difference between the chains is not a conserved quantum number, and hence the corresponding $U(1)$ symmetry is broken. A consequence of this is that $\mathcal{F}^-(\ell)$ and $\mathcal{F}^-(L-\ell)$ are no longer supposed to be equal; this is most easily illustrated by the relation $\mathcal{F}^-(L) \neq \mathcal{F}^-(0) = 0$. These considerations allow for linear contributions\cite{herviou_et_al_2016} to the fluctuations of the relative particle number:
\begin{equation}
  \label{Eq:FmellMOBC}
  \mathcal{F}_\textit{Meissner}^-(\ell) = b \ell + c + d b_E(\ell),
\end{equation}
where the coefficient of the linear term is $b \propto 1/\xi^{-}$, with $\xi^{-}$ the correlation length of~(\ref{Eq:RR}). This is because $\mathcal{F}^{-}$ is the leading contribution to $\log \langle e^{i \sqrt{2} \phi^-(0)} e^{-i \sqrt{2} \phi^-(\ell)} \rangle \sim -\ell/\xi^{-} + \textit{const}$. Note that in the above expression it is the length of the subsystem $\ell$ that enters, and not the compactified distance $d(\ell|L)$. Secondly, total charge fluctuations in the Meissner phase $\mathcal{F}^+_\textit{Meissner}(\ell)$ are of the form $\mathcal{F}^+_\textit{vortex}(\ell)$ in Eq.~(\ref{Eq:FpmellOBC}), on account of the fact that the total charge field is gapless in both phases.

\begin{figure}[t!]
  \includegraphics[width=\linewidth]{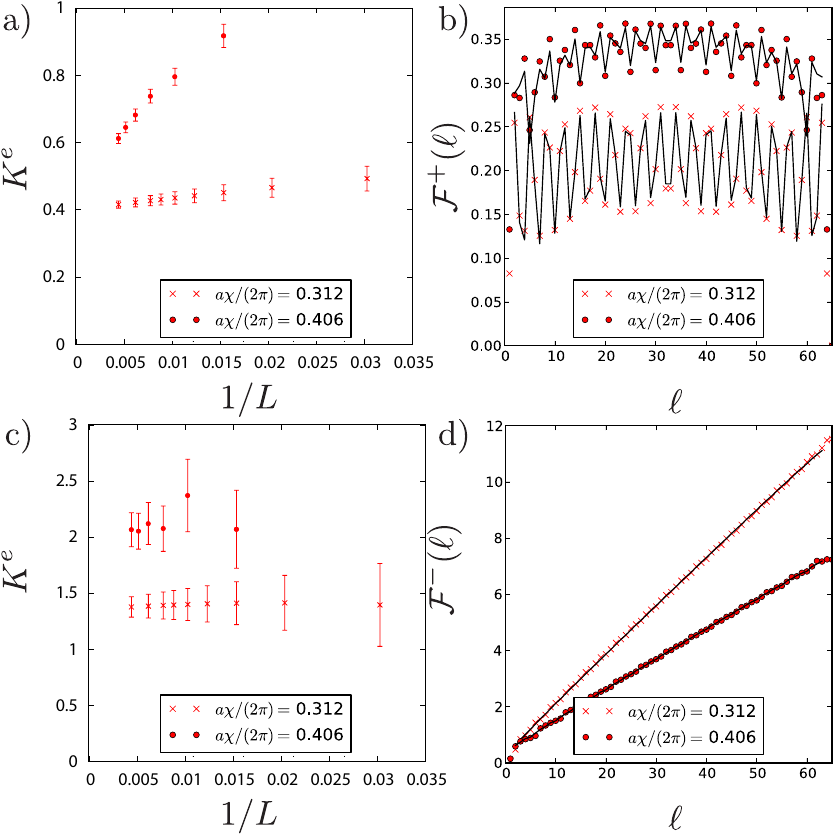}
  \caption{\label{Fig:FPL}(Color online) Fit results for edge Luttinger parameter, $K^e$, from particle number fluctuations for two points on the phase diagram of Fig.~\ref{Fig:P}, $a n_0 = a\chi/(2\pi)=0.3125$ (red crosses) and $0.406$ (red dots): a) Values of $K^e$ extracted from the fit of $\mathcal{F}^+(\ell)$ according to Eq.~(\ref{Eq:FtFPLOBC}), for $L \in [33,225]$, at filling factor $\nu=1/2$. Extrapolation to $L\to \infty$ is consistent with $K^e=0.4$. b) Fit function (black curve) versus numerical data [same symbols as in a)] according to Eq.~(\ref{Eq:FtFPLOBC}) for $L = 65$. c) Values of $K^e$ extracted from the fit of $\mathcal{F}^-(\ell)$ according to Eq.~(\ref{Eq:FmellLOBC}), for $L \in [33,225]$, at fixed filling factor $\nu=1/2$. d) Fit function (black curve) versus numerical data according to Eq.~(\ref{Eq:FmellLOBC}) for $L = 65$. The linear component to the $\mathcal{F}^{-}(\ell)$ is dominant in the Laughlin phase, which makes the fit for the logarithmic contribution difficult. We attribute to this the larger error bars in panel c).}
\end{figure}

In the Laughlin phase, we reexpress density fields according to our definitions in Sec.~\ref{Sec:M} and App.~\ref{App:ET}
\begin{eqnarray}
  \sqrt{2}\phi^- = \theta' - \theta, \; \sqrt{2}\phi^+ = \phi + \phi',
\end{eqnarray}
where $\theta'$ and $\phi'$ correspond to the edge fields. Therefore we use as a template function
\begin{equation}
  \label{Eq:FmellLOBC}
  \mathcal{F}_\textit{Laughlin}^-(\ell) = \frac{1}{2\pi^2 K^e} \log\left[ d(\ell|L) \right] + b \ell + c + d b_E(\ell),
\end{equation}
where the linear contribution is prefactored by the parameter $b$, which scales like the inverse correlation length associated with the Laughlin gap $\Delta$ in Eq.~(\ref{Eq:LaughlinGap}). Total density fluctuations are
\begin{equation}
  \label{Eq:FtFPLOBC}
  \mathcal{F}_\textit{Laughlin}^+(\ell) = \frac{K^e}{2\pi^2} \log\left[ d(\ell|L) \right] + c + d b_E (\ell).
\end{equation}
Figure~\ref{Fig:FPL} shows results for $K^e$ as extracted from fits to $\mathcal{F}^\pm(\ell)$ obtained in systems with lengths $L \in [33,225]$ with boson numbers $N=0.625(L-1)/2$ and $0.8125(L-1)/2$ at fixed filling fraction $\nu=1/2$. The $K^e$ obtained from fits of $\mathcal{F}^+$ is in agreement with the theoretical prediction $K^e = 0.4$ (see Sec.~\ref{Sec:B} and App.~\ref{App:ET}). The values obtained from fits of $\mathcal{F}^-$ have larger error bars and do not extrapolate to the value determined from the fit of $\mathcal{F}^{+}$. We attribute this to an overall worse quality of the fit in the presence of the dominating linear contribution.

In conclusion, bipartite charge fluctuations probe the correlations of gapless fields and access the edge Luttinger parameter $K^e$. There is quantitative agreement between DMRG results and the edge theory of Sec.~\ref{Sec:B}. The Josephson gap can be indirectly extracted from the linear contribution to $\mathcal{F}^-(\ell)$ in Eq.~(\ref{Eq:FmellMOBC}) [The Laughlin gap is obtained, analogously, from Eq.~(\ref{Eq:FmellLOBC})]. For example, with this method one can confirm the power law dependences of Eqs.~(\ref{Eq:Deltam}) and~(\ref{Eq:LaughlinGap}). This computation, not pursued here, remains the object of future work.

\subsection{Correlation functions}
\label{Subsec:C}
In this subsection, we extract exponentially decaying contributions in correlation functions as signatures of the formation of spectral gaps. We find that this procedure is straightforward in the Meissner phase and forms a hallmark of the Josephson gap $\Delta^-$ of Eq.~(\ref{Eq:Deltam}). The exponential behavior distinguishes the Meissner phase from the Laughlin and vortex phases, where correlation functions associated to the Josephson phase $\theta^-$ vanish algebraically. In fact, we find that near the commensurate line $\chi/(2\pi)=n_0$ the Josephson gap is weakened -- in the sense that exponential decay of the Josephson correlation function is suppressed.

Second, we attempt in this subsection the construction of DMRG accessible correlation functions to capture the exponential decay of the gapped field $\phi$ corresponding to the Laughlin phase. We discuss the drawback of this in comparison to the charge fluctuations discussed above.

\begin{figure}[t!]
  \includegraphics[width=\linewidth]{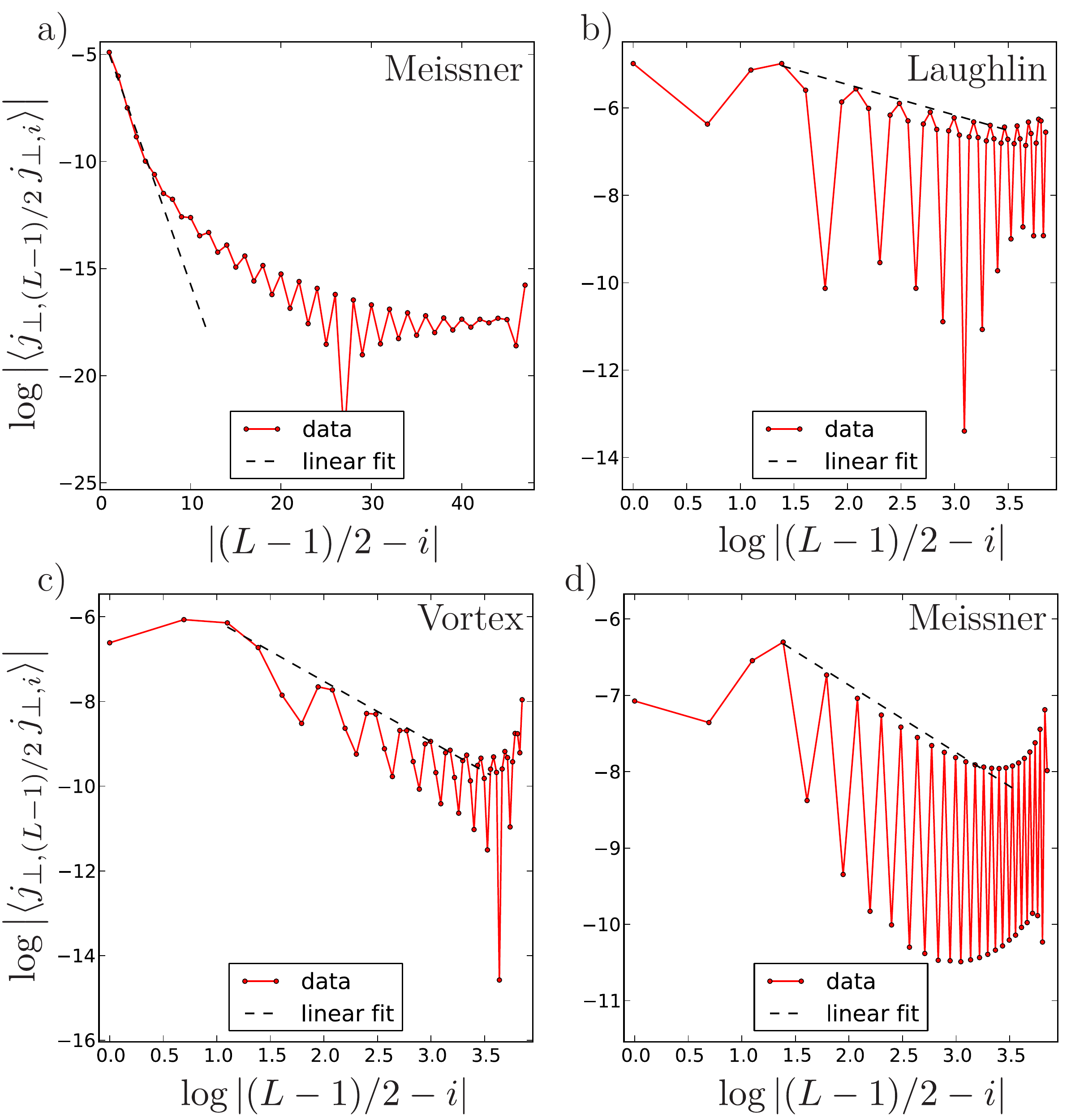}
  \caption{\label{Fig:RRlili} (Color online) Rung current two point correlation function, $\langle j_{\perp,(L-1)/2} \, j_{\perp,i} \rangle$ for $L=97$ rungs with OBC, decays a) exponentially in the Meissner phase, at $a\chi/(2\pi)=0.125,\, an_0=0.5$; b) algebraically in the Laughlin phase, $a\chi/(2\pi)=an_0=0.375$; c) algebraically in the vortex phase, $a\chi/(2\pi)=0.375, an_0 = 0.25$. d) In the Meissner phase at $\chi/(2\pi)=n_0$ correlations can be algebraic, e.g. at $a\chi/(2\pi)=an_0=0.25$. We use log-linear (log-log) coordinates for exponential (algebraic) decay, to fit linear templates, which we shift by a constant to guide the eye.}   
\end{figure}

The condensation of the Josephson phase yields exponentially decaying rung current correlation functions, as in Eq.~(\ref{Eq:RR}). We calculate the correlator $\langle j_\perp(0) j_\perp(x) \rangle$ between a boundary rung and a rung situated at position $x=ia$ in open boundary conditions (other choices for the two points of the correlator give qualitatively similar results). This correlator  decays exponentially in the Meissner phase at high density and low flux [see Fig.~\ref{Fig:RRlili}]. There is algebraic decay to a constant amplitude oscillation at large distance in the Laughlin phase, which indicates that the Josephson term is no longer responsible for the gapped mode. In fact, for $2\pi n_0 = \chi$ but at low enough flux deep in the Meissner phase the exponential character of the correlator is suppressed, suggesting a competition between two different gapping mechanisms. Finally, the vortex phase has algebraically decaying correlations, as expected from bosonization. In conclusion, there are $c=1$ regions in which the Josephson correlation function is not exponentially decaying. This is consistent with the existence of a gapped field which is distinct from $\sqrt{2}\theta^-(x)$ in Sec.~\ref{Sec:B}. 

Secondly, note that the Laughlin vertex operator $\phi$ is present in correlation functions of tight binding operators. One such example is the string operator corresponding to total density,
\begin{equation}
  s_{+,i_0} \equiv e^{i \alpha \pi (N_{1,i_0} + N_{2,i_0})},
\end{equation}
where $\alpha$ is a real parameter. String operators have been used previously to determine the ``Haldane insulator'' phase of bosons in one dimensional lattices with long range repulsion \cite{dalla_torre_et_al_2006, berg_et_al_2008}.  The string operator considered here can be expressed in bosonized form $s_{+,i_0} \to s_{+}(x_0)$ in terms of the fields $\phi$ and $\phi'$:
\begin{eqnarray}
 s_{+}(x_0) &=& e^{i \alpha \pi( n_0^1 + n_0^2 )x_0 - i \alpha \sqrt{2} \phi^+(x_0) + i \alpha \sqrt{2} \phi^+(0) }  \\
 &=& e^{i \alpha \pi( n_0^1 + n_0^2 )x_0 - i \alpha [ \phi(x_0) + \phi'(x_0)] + i \alpha [ \phi(0) + \phi'(0)] }. \nonumber
\end{eqnarray}
Two--point correlation functions of $S_+$ take the form
\begin{eqnarray}
  \label{Eq:SplusSplusCorrelator}
  \langle s_{+}(x) s_{+}^{\dagger}(x') \rangle &=& e^{i \alpha \pi (n_0^1 + n_0^2 ) (x - x') }  \times 
  \\
  && \;\;\;\;\;\; \left\langle e^{-i \alpha \left[ \phi(x) - \phi(x') \right]}  
  e^{-i \alpha \left[ \phi'(x) - \phi'(x') \right]} \right\rangle,  \nonumber 
\end{eqnarray}
where we have disentangled the exponentials since all involved field operators commute. Consider the spatial dependence of this correlation function in the Laughlin phase, where the bulk field $\phi$ is gapped and $\phi'$ is a gapless field corresponding to the edge Hamiltonian $\mathcal{H}^e$ in Eq.~(\ref{Eq:Hec}). The expectation value separates into a product of expectation values,
\begin{eqnarray}
  \left\langle e^{-i \alpha \left[ \phi(x) - \phi(x') \right]}  \right\rangle &\sim& \text{const.} + \text{const.} \times e^{-|x-x'|/\xi},
  \\
  \left\langle e^{-i \alpha \left[ \phi'(x) - \phi'(x') \right]}  \right\rangle &\sim& \left|\frac{1}{x-x'}\right|^{\alpha^2 K^e},
\end{eqnarray}
where in the first $\xi \sim v / \Delta$ is the correlation length of the Laughlin gap, and the correlator decays exponentially because the field $\phi$ is condensed; and in the second, the power law decay comes from the Gaussian edge theory with Luttinger parameter $K^e$, Eq.~(\ref{Eq:He}). A numerical study of these correlation functions shows manifest power law decays, which indicates that the correlation length $\xi$ is very large, or equivalently that the gap $\Delta$ is small.

We end the section on DMRG with a final remark on the role of rung repulsive interactions. In the dilute limit considered here $a n_0 < 0.5$, moderate values of $V_\perp$ of up to a few $t$ did not show significant differences from $V_\perp = 0$. Although the perturbative bosonization analysis of Sec.~\ref{Sec:B} would suggest the contrary, it is possible that the pinning potential of Eq.~(\ref{Eq:LaughlinCouplingm}) is relevant even at $V_\perp=0$ in the dilute limit. This is a hypothesis, as we have been unable to extract the scaling dimension of Eq.~(\ref{Eq:SDL}) from correlation functions.

\section{Thin torus geometry}
\label{Sec:D}
This section deals with the evaluation of ground state observables and the Hall conductivity for geometry which is modified compared to Fig.~\ref{Fig:L}a), in which the boundaries in both $x$ and $y$ directions are closed so as to realize a thin torus. The studies in this section rely on the exact diagonalization \cite{bauer_et_al_2011} of small lattices, and aim to highlight topological properties of the bulk when the edge channels are gapped out. The Hamiltonian for an $L \times 2$ site torus [see Fig.~\ref{Fig:L}b)] whose small perimeter is threaded by an Aharonov--Bohm phase $\theta_y$ and large perimeter is threaded by $\theta_x$ is
\begin{eqnarray}
\label{Eq:H_thetay}
H(\theta_y) &=& -g \sum_{j=1}^{L} \left[ b^\dagger_{2,j} b_{1,j} e^{i\left( a \chi j + \frac{\theta_y}{2} \right)} + \text{H.c.}\right] \nonumber \\
 && -g' \sum_{j=1}^{L} \left[ b^\dagger_{1,j} b_{2,j} e^{i\left( a \chi j + \frac{\theta_y}{2} \right)} + \text{H.c.} \right] \nonumber \\
 && -t \sum_{j=1}^{L} \left[ e^{i\frac{\theta_x}{L}} b^\dagger_{1,j+1} b_{1,j} +e^{i\frac{\theta_x}{L}} b^\dagger_{2,j+1}b_{2,j} +\text{H.c.} \right] \nonumber \\
 && - \sum_{j=1}^{L} \left( \mu_{1,j} b^\dagger_{1,j} b_{1,j} + \mu_{2,j} b^\dagger_{2,j} b_{2,j}  \right).
\end{eqnarray}
Note that if $g' = 0$ and $\theta_{x,y}=0$, we recover~(\ref{Eq:H}) up to a gauge transformation, while if $g'=g$ is the thin torus case analyzed by Grusdt and H\"oning \cite{grusdt_hoening_2014}. The torus spectrum no longer has a gapless edge channel; it is gapped by the bond $g'$ closing the periodic boundary in the $y$ direction. With this geometry one can probe the bulk polarization by removing the effect of the gapless edge. Unless otherwise stated, we will be concerned in this section with the case
\begin{equation}
g'=g.
\end{equation}
Note also that the last row of Eq.~(\ref{Eq:H_thetay}) contains chemical potential terms, used below to split the degeneracy of the ground state manifold. 

\subsection{Ground state degeneracy} 
\label{Subsec:GSD}
In this subsection, we first argue that the ground state is twofold degenerate and that the effect of threading one flux quantum [varying $\theta_y$ between $0$ and $2\pi$] is to interchange the two ground states, in agreement with the theory of ground state manifold degeneracies for abelian fractional quantum Hall states \cite{thouless_1989}. We confirm this numerically using exact diagonalization \cite{bauer_et_al_2011}. The arguments in this subsection follow from a more general case discussed by Oshikawa\cite{oshikawa_2000}.

Assume that $H(\theta_y)$ has two lowest lying eigenstates, denoted ``0'' and ``1'', which we express using the following notation
\begin{eqnarray}
\label{Eq:EigenSysNot}
H( \theta_y ) | \psi_0( \theta_y ) \rangle &=& E_0( \theta_y ) | \psi_0 ( \theta_y ) \rangle, \nonumber \\
H( \theta_y ) | \psi_1( \theta_y ) \rangle &=& E_1( \theta_y ) | \psi_1 ( \theta_y ) \rangle.
\end{eqnarray}
We convene that labels ``0'' and ``1'' are assigned to the eigenstates such that ``0'' is always the state of lower energy, that is
\begin{equation}
\label{Eq:EnConv}
E_0( \theta_y ) \leq E_1( \theta_y ).
\end{equation}
With such a choice, in the presence of level crossings, where a relabeling is necessary, the energies are not smooth functions of $\theta_y$. We argue here that there is at least one level crossing for $\theta_y$ between $0$ and $2\pi$, equivalently that there exists some $\theta_y$ such that there is an equality in Eq.~(\ref{Eq:EnConv}). We now summarize the argument, saving details for App.~\ref{App:GSD}.

Assume that the two lowest states of $H(0)$ are nondegenerate. If they are degenerate, the degeneracy can be split by adding infinitesimal $\mu_{i,j}$ that breaks translation symmetry in the $x$ direction but not in the $y$ direction. The two lowest eigenstates of $H(2\pi)$ obey for each $\alpha=1,2$
\begin{equation}
  \label{Eq:EnergyEquality}
  E_\alpha (0) = E_\alpha (2\pi).
\end{equation}
This follows from the fact that $H(0)$ and $H(2\pi)$ are related by a gauge transformation
\begin{equation}
\label{Eq:H_gtransform}
U_y^\dagger( - 2 \pi ) H( 2 \pi ) U_y( - 2 \pi ) = H( 0 ),
\end{equation}
realized by the following unitary operator
\begin{equation}
U_y^\dagger(\theta_y) \equiv \exp\left( i \sum_{i,j} j n_{j,i} \frac{\theta_y}{L_y} \right),
\end{equation}
with $L_y=2$. The spectrum of $H(\theta_y)$ returns to itself after a full flux quantum has been threaded.

A level crossing of $E_0(\theta_y)=E_1(\theta_y)$ for some $\theta_y$ is necessary whenever the $y$ momentum quantum number associated with the eigenstates at $\theta_y=0$ and $\theta_y=2\pi$ changes. To show when this is the case, we use Eq.~(\ref{Eq:H_gtransform}) to find a relation between the eigenvectors at $\theta_y=0$ and $2\pi$:
\begin{equation}
|\psi_\alpha (0) \rangle = U_y(2\pi) |\psi_\alpha(2\pi) \rangle.
\end{equation}
Note, however, that $|\psi_\alpha(0) \rangle$ and $|\psi_\alpha(2\pi) \rangle$ may be distinct, since they may correspond to distinct eigenvalues of the $y$ momentum, 
\begin{eqnarray}
P_y = \sum_{k_x,k_y} b_{k_y,k_x}^\dagger b_{k_y,k_x} k_y,
\end{eqnarray}
which is a good quantum number by virtue of the $y$-translation invariance and whose eigenvalues, in units of the inverse lattice constant, are $0$ and $\pi$, for $L_y=2$. Whenever $E_{0}(\theta_y) \neq E_{1}(\theta_y)$, the two ground states $|\psi_\alpha(\theta_y)\rangle$ are eigenstates of momentum, with eigenvalues $P_{y,\alpha}(\theta_y)$. These eigenvalues obey
\begin{equation}
  P_{y,\alpha}(0) = P_{y,\alpha}(2\pi) - N \pi,
\end{equation}
which implies that $P_y$ changes as one flux quantum is threaded at odd particle number $N$. This change implies there has been at least one crossing between the two levels.

\begin{figure}
  \includegraphics[width=\linewidth]{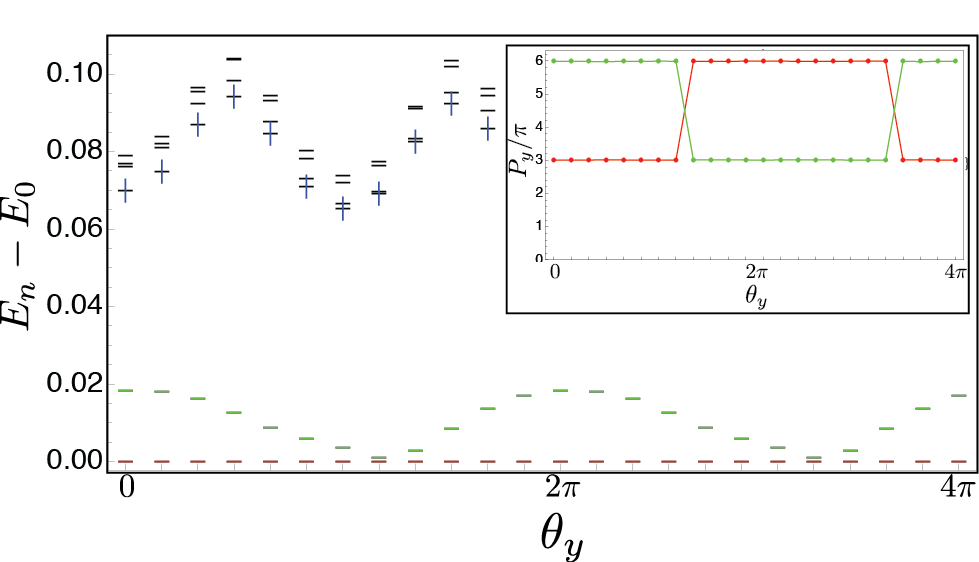}
  \caption{\label{Fig:SF} (Color online) Low-lying spectrum $E_n - E_0$ for $n \geq 0$, versus $\theta_y$, as obtained from the Jacobi--Davidson diagonalization\cite{bauer_et_al_2011} of 3 bosons on a 12 rung thin torus at $\nu = 1/2$ (see text for details). The ground state manifold (corresponding to $E_0$ and $E_1$) is highlighted with red, green symbols, and excited states in blue ($E_2$) and black ($E_n$ for $n \geq 3$). The inset shows $P_{y,n}(\theta_y)$ for the lowest states $n=0,1$: The quantum numbers $P_y$ of the ground states are interchanged at the level crossings, and there are two crossings in the interval $\theta_y \in [0,4\pi)$.}
\end{figure}

To confirm the above, we use exact diagonalization to find the ground and first few excited states of $H(\theta_y)$ for $\theta_y \in [0,2\pi)$ for a torus of length $L = 12$, flux per square plaquette $a \chi = 3 / 12 \times 2\pi$ and $N=3$ bosons, corresponding to $\nu = 1/2$. The magnetic unit cell consists of $4$ plaquettes. There are 3 magnetic unit cells in the $x$ direction, and 2 magnetic unit cells in the $y$ direction of the torus. In this low--dimensional realization, the ground state multiplet is composed of two charge density waves \cite{grusdt_hoening_2014}. As $\theta_y$ traverses $[0,2\pi)$, the energies of two quasi--degenerate ground states are interchanged, and return to their original values after $\theta_y$ traverses another period from $[2\pi, 4\pi)$ (see Fig.~\ref{Fig:SF}). At the degeneracy point, the quantum numbers $P_y$ (there are two possibilities for the eigenvalues of $P_y$, 0 and $\pi/a' \text{ mod } 2\pi/a'$) are interchanged between the two ground states (cf. inset of Fig.~\ref{Fig:SF}).  

\subsection{Thouless pump and Hall conductivity}
The Hamiltonian $H(\theta_y)$ realizes a fractional 1/2 charge pump \cite{thouless_1983,grusdt_hoening_2014}. Equivalently, the shift of the center of mass of each of the two ground states is half of the magnetic unit cell.

We evaluated the Hall conductivity $\sigma_{xy}$ in the ground state multiplet and defined in terms of twisted boundary conditions \cite{niu_et_al_1984}:
\begin{equation}
  \sigma_{xy} = \frac{1}{d} \frac{1}{2\pi i} \int d^2\theta \text{Tr}  \left[ \langle \partial_{\theta_x} \Psi | \partial_{\theta_y} \Psi \rangle - \langle \partial_{\theta_y} \Psi | \partial_{\theta_x} \Psi \rangle \right].
  \label{Eq:CNWTN}
\end{equation}
$d$ is the ground state multiplet degeneracy. $|\Psi\rangle$ is the ground state multiplet. In the numerical evaluation, it is necessary to fix the gauge on the eigenvectors of the 2-fold degenerate ground state multiplet following a method of Hatsugai \cite{hatsugai_2005,*hatsugai_2004}, which is a generalization of the winding number argument put forth by Kohmoto \cite{kohmoto_1985}. For $\nu=1/2$, 4 particles on 12 rungs, at flux $a\chi = \pi/2$ per plaquette, the resulting $\sigma_{xy} = 0.573$; we attribute the discrepancy to finite size effects.

\begin{figure}[t!]
  \includegraphics[width=0.5\linewidth]{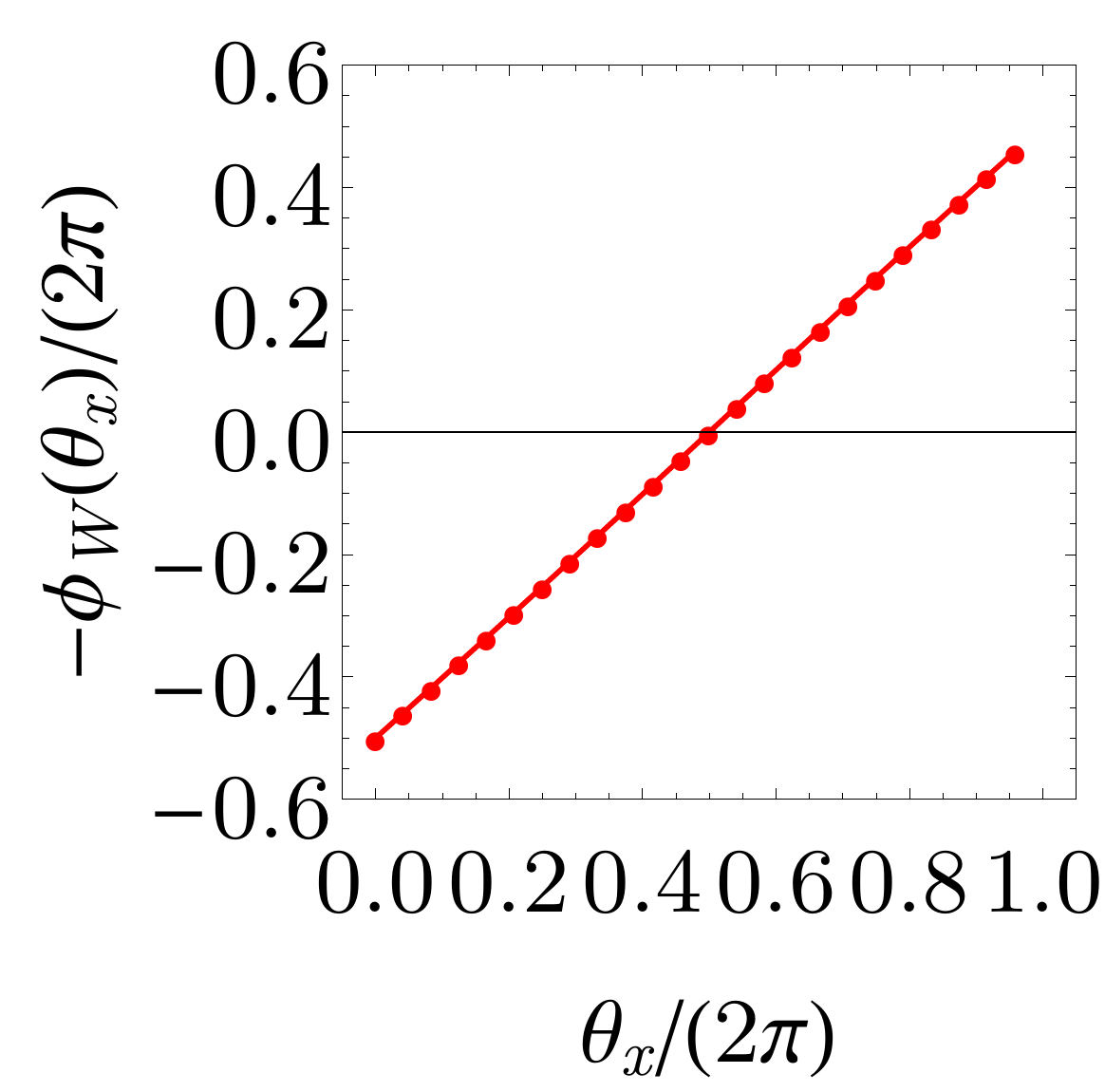}
  \caption{\label{Fig:WX}(Color online) Many body Zak phase $\phi_W(\theta_x)$ as computed from the Wilson loop advances by $2\pi$ when $\theta_x \in [0,2\pi]$. Twists $(\theta_x,\theta_y) \in [0,2\pi]^2$ discretized on $24 \times 12$ mesh.}
\end{figure}

The expression in Eq.~(\ref{Eq:CNWTN}) can be computed without the need to gauge fix as follows\cite{yu_et_al_2011}. With the ground state manifold denoted by $\{| m, \theta_x,\theta_y\rangle | m = 1,2 \}$ at $\theta_x,\theta_y$, let
\begin{eqnarray}
  W^{mn}(\theta_x) = \langle m, \theta_x,\theta_{y,0}| n_1, \theta_x,\theta_{y,1} \rangle \nonumber \times \\
\langle n_1, \theta_x,\theta_{y,1} | n_2, \theta_x,\theta_{y,2}  \rangle \times ... \nonumber \\ 
\langle n_{N_y - 1}, \theta_x, \theta_{y,N_y - 1} | n, \theta_x,\theta_{y,0} \rangle,
\end{eqnarray}
where $m, n = 1,2$ are indices for states in the ground state manifold, and summation over $n_1,...,n_{N_y-1}=1,2$ is implicit. Let $\theta_{y,j} = \theta_{y,0} + j \Delta \theta_y$, where $\Delta \theta_y$ is the discretization step. Then $W^{mn}(\theta_x)$ is the Wilson loop matrix for twist angle $\theta_x$. The argument of its determinant is the many body generalization of the single particle Zak phase \cite{zak_1989} for a ground state doublet:
\begin{equation}
  \label{Eq:WX}
  \phi_W(\theta_x) = \text{Im} \log \text{Det} W(\theta_x),
\end{equation}
is plotted in Fig.~\ref{Fig:WX}, while $\sigma_{xy}$ is related to the winding of this phase:
\begin{eqnarray}
  \sigma_{xy} &=& \frac{1}{d} \frac{1}{2\pi} \int_0^{2\pi} d\theta_x \frac{\partial}{\partial \theta_x} \phi_W(\theta_x) \nonumber \\ 
  &=& \frac{1}{d} \frac{1}{2\pi} \left[ \phi_W(\theta_{x,N_x}) - \phi_W(\theta_{x,0}) \right].
\end{eqnarray}
If $\Phi_W(\theta_x)$ changes by $2\pi$ upon traversal of $\theta_x \in [0,2\pi)$, then $\sigma_{xy}=1/2$.

Finally, another way to interpret $\sigma_{xy}=1/2$ is that the thin torus system realizes a fractional Thouless pump. Upon threading a flux quantum along the short perimeter of the thin torus, the two ground states, which are displaced by half of the magnetic unit cell, are interchanged. Upon threading another flux quantum, a second interchange of the states occurs, and the two charge density waves return to the original locations (modulo translations by the size of the magnetic unit cell). This is consistent with the interpretation of the Zak phase in terms of the center of mass coordinate of the many body state, which is a generalization of the relation between the Zak phase of a Bloch band and the band center \cite{zak_1989}.

In low dimensional geometries, a formulation of the bulk--edge correspondence in terms of Zak phases can be used to predict the number of intragap edge states\cite{rhim_et_al_2016}. Recent experiments with condensates of weakly interacting rubidium atoms in dimerized lattices have allowed one to measure the Zak phase \cite{atala_et_al_2013} (using a combination of Ramsey interferometry and Bloch oscillations, which can be generalized to 2D and 3D lattices\cite{grusdt_et_al_2014}) as well as a quantized Thouless pump via \textit{in situ} imaging of the atomic cloud \cite{lohse_et_al_2015}. The Thouless pump was demonstrated as well with the fermionic species $^{171}$Yb \cite{nakajima_et_al_2015}.

\section{Conclusions}
\label{Sec:C}
We have studied the phase diagram of a two-leg ladder of hard core bosons in uniform flux, as a function of boson density and flux, focusing on the dilute limit of up to 1/2 bosons per unit cell. We uncovered signatures of a commensurability effect between flux and density, associated with the emergence of the precursor Laughlin phase. We propose that this phase diagram contains three regions: the Meissner phase, characterized by Josephson phase coherence, the vortex phase, a gapless phase in which all correlations are algebraic, and the Laughlin phase. The Laughlin phase is distinct from the neighboring Meissner phase (both of these phases have central charge $c=1$), through vortex patterns in expectation values of local bond current operators.  The existence of Meissner phases and Vortex phases was previously proven numerically \cite{piraud_et_al_2014}. Here, we present numerical evidence supporting the existence of a Laughlin phase. For the chosen value of interchain hopping and dilute limit, we find that the rung interaction $V_{\perp}$ does not seem important to see the phase. 

We note related recent work on bosonic ladders. In Ref.~\onlinecite{greschner_et_al_2016}, an incommensurate Meissner phase has been observed numerically for density $a n_0=0.4$ at flux $a\chi/(2\pi) = 0.4$, which belongs to the precursor-to-Laughlin state region on our phase diagram. The study in Ref.~\onlinecite{calvanese_strinati_et_al_2016} explores Laughlin-like states at $\nu=1/2$ in a different parameter regime of $g \sim 10^{-2} t$ and $V_\perp \neq 0$. A comparison of this study with our own is the subject of future work.  References~\onlinecite{di_dio_et_al_2015}~and~\onlinecite{orignac_et_al_2016} discuss the appearance of an incommensuration in the vortex state when the flux per plaquette is $\chi = \pi n_0$, where $n_0$ is the boson filling per rung. That parameter choice is nevertheless distinct from the one chosen in this paper, $\chi = 2 \pi n_0$.

Unlike in the case of coupled wire constructions of quantum Hall states \cite{teo_kane_2014, kane_et_al_2002, neupert_et_al_2014,*huang_et_al_2016}, where edge backscattering terms are exponentially small in the large number of chains, the quasi--one--dimensional system presented has a gapless edge channel which contains backscattering terms. Therefore, $K^e = 1/2$, corresponding to a Laughlin state at filling factor $\nu=1/2$, can only be obtained by fine tuning long--range repulsive interactions. For hard core bosons without repulsive interactions, we have calculated $K^e=2/5$ from a bosonized theory, a value compatible with the DMRG calculations. We have argued that since $K^e$ is related to charge fractionalization at the edge, it can be observed in scattering experiments, such as the ones performed for 2D electron gases \cite{de_picciotto_et_al_1997,*radu_et_al_2008}, which are feasible with ultracold atom quantum point contacts \cite{brantut_et_al_2012}.

To obtain $K^{e}$, we have used bipartite particle number fluctuations, in which this coefficient is a prefactor of the logarithmic contribution. Specifically, we have distinguished between symmetric and antisymmetric bipartite fluctuations to account for the presence of a gapped mode, which produces a dominant linear contribution in the antisymmetric sector. $K^{e}$ is more accurately extracted from the symmetric bipartite fluctuations. More generally, this illustrates that bipartite charge fluctuations can sharply detect properties of topological phases by furnishing quantitative probes of their gapless excitations. In free fermion systems, bipartite number fluctuations are related to entanglement entropies or the entanglement spectrum\cite{song_et_al_2010,song_et_al_2011,petrescu_et_al_2014}. Moreover, we have shown that results obtained from bipartite fluctuations are consistent with the algebraic or exponential decays of a selection of correlation functions. In this sense, bipartite fluctuations, which can be interpreted as logarithms of exponential correlators, are the more easily accessible, both numerically in DMRG and experimentally.

We have also provided probes to measure fractional charges at the edge (for a discussion of fractionally charged quasiparticles, see App.~\ref{App:ET}) and the bulk polarization.  Even though the charges at the edges would be sensitive to small backscattering effects, a fractional charge $\nu=1/2$ could be potentially inferred from the polarization in a transverse Hall conductivity measurement on a torus. Moreover, we have shown here that the Laughlin state is characterized by a line of local extrema in the expectation value of the antsymmetric parallel current $\langle j_\parallel^{-}\rangle$. This observable is readily available in experiments.

Our starting point model in Eq.~(\ref{Eq:H}) is realizable with recently developed experimental capabilities. For example, Floquet protocols\cite{goldman_dalibard_2014, cayssol_et_al_2013} with $^{87}$Rb atoms via laser assisted tunneling in square optical lattices \cite{aidelsburger_et_al_2014, miyake_et_al_2013, atala_et_al_2014}. Two of us have argued \cite{petrescu_le_hur_2015} that the model in~(\ref{Eq:H}) can be realized near the Tonks gas limit of hard core bosons by mapping the leg index $1,2$ to one of two hyperfine states of $^{87}$Rb. In this fashion, multileg ladders have been obtained \cite{celi_et_al_2014,cooper_rey_2015,pagano_et_al_2014,mancini_et_al_2015, stuhl_et_al_2015}.

We end with another note on the experimental realization. The Peierls phase factors required to realize the model studied in this work can be obtained by modulating the lattice in time, based on protocols introduced for photonic resonator lattices \cite{le_hur_et_al_2016}. We illustrate this by considering a pair of sites with time dependent tunneling amplitude~\cite{fang_et_al_2013}
\begin{eqnarray}
  H(t) &=& H_0 + H_\textit{tunnel}(t) + H_{\textit{int}},
  \nonumber \\
  H_0 &=& \frac{\omega_A}{2} \sigma_A^z + \frac{\omega_B}{2} \sigma_{B}^z, 
  \nonumber \\
  H_\textit{tunnel}(t) &=& V \cos( \Omega t + \phi_{AB} ) \left( \sigma_A^+ \sigma_B^- + \text{H.c.} \right).
\end{eqnarray} 
Any intersite interaction is encoded in $H_{\textit{int}}( \sigma_A^z , \sigma_B^z )$. Assume that $\omega_A < \omega_B$, and that the modulating frequency is resonant with the energy difference of the two site $\Omega = \omega_B - \omega_A$. Note the nonzero phase $\phi_{AB}$ of the drive. 

Consider switching to an interaction picture with respect to $H_0$, \textit{i.e.} let $H' = U(t) [H(t) - i \partial_t] U^{\dagger}(t)$, with $U(t) = \exp( i H_0 t)$, leading to 
\begin{equation}
  H' = V \cos( \Omega t  + \phi_{AB} ) \left( \sigma_{A}^{+} \sigma_{B}^{-} e^{- i  \Omega t } + \text{H.c.} \right) + H_{\textit{int}} .
\end{equation}
Assume that $V \ll \Omega$ in order to perform the rotating wave approximation, amounting to dropping the terms oscillating at $2\Omega$, the above becomes
\begin{equation}
  H'_{\text{RWA}} = V \left( \sigma_{A}^{+} \sigma_{B}^{-} e^{ i \phi_{AB} } + \sigma_{B}^{+} \sigma_{A}^{-} e^{- i \phi_{AB} }  \right) + H_{\textit{int}}.
\end{equation}
In the rotating wave approximation and provided that the drive is resonant with the frequency detuning between the sites, the eigenvalues of $H(t) - i \partial_t$, called quasienergies\cite{sambe_1973},  correspond to the eigenstates of $H'_{\text{RWA}}$ \cite{fang_et_al_2013}. The interaction term is unaffected by the transformations that lead to the effective time--independent Hamiltonian. This is an important difference between the hard core bosons discussed here and Bose--Hubbard models, where density assisted hopping terms are possible in the Magnus expansion\cite{plekhanov_et_al_2016}.

In Eq.~(\ref{Eq:H}), a sensible choice is to put all Peierls phases on the horizontal bonds, $A_{\perp,i}=0$, $A_{i,i+1}^2 = -\chi$, and $A_{i,i+1}^1 = 0$, together with imposing a gradient in the onsite energies of the second chain $\sum_{i=1}^{L} (\Omega \times i) b^\dagger_{2,i} b_{2,i}$. The thin torus geometry can be achieved by a simple replacement $V \cos( \Omega t  + \phi_{AB} ) \to V \cos( \Omega t  + \phi_{AB} ) + V' \cos( \Omega t  - \phi_{AB} )$, with an analogous analysis. Note that the model discussed here, which in the absence of long ranged interactions is an $XY$ exchange Hamiltonian, can be obtained in arrays of coupled QED cavities operated in the photon blockaded regime\cite{angelakis_et_al_2007}.

\section*{Acknowledgments}
We are grateful to Monika Aidelsburger, Immanuel Bloch, Fabian Heidrich-Meisner, Thierry Giamarchi, Sebastian Greschner, Lo\"ic Herviou, Walter Hofstetter, Leonardo Mazza, Christophe Mora, Edmond Orignac, Guido Pagano, Bel\'{e}n Paredes, Kirill Plekhanov, Antoine Sterdyniak, Leticia Tarruell, and Temo Vekua for helpful discussions. A.P. and K.L.H. acknowledge CIFAR for workshops where parts of this work have been presented. K.L.H. would also like to thank KITP Santa Barbara for hospitality. M.P. was supported by the European Union through the Marie-Curie grant ToPOL (No.~624033) (funded within FP7-MC-IEF). I.P.M. acknowledges support from the Australian Research Council (ARC) Centre of Excellence for Engineered Quantum Systems, grant CE110001013, and the ARC Future Fellowships scheme, FT140100625. K.L.H. acknowledges support from the German DFG Forschergruppe 2014 and Labex PALM.

\appendix
\section{Bosonizing hard core bosons}
\label{App:EquivB}
In this appendix we provide a bosonization scheme equivalent to the one in the text, which entails passing from hard core bosons to spin--$1/2$ moments and then to fermions via a Jordan--Wigner transformation.
\subsection{Bosonized Hamiltonian for a single chain}
We recapitulate the bosonized Hamiltonian for a single chain \cite{giamarchi_2003} of spin--$1/2$ moments
\begin{equation}
H = \sum_i J_{xy} (S_{i+1}^x S_i^x + S_{i+1}^y S_i^y) + J_z S_{i+1}^z S_i^z.
\end{equation}
The continuum Hamiltonian describing the low-energy degrees of freedom is expressed in terms of canonically conjugate fields $[\phi(x),\theta(x')] = i \frac{\pi}{2} \text{Sign}(x'-x)$. It takes the form
\begin{eqnarray}
H &=& H_0 - \frac{J_z}{2\pi^2 a} \int dx \cos( 4 \phi ), \nonumber \\
H_0 &=& \frac{v}{2\pi} \int dx \left[ K(\nabla \theta)^2  + \frac{1}{K}(\nabla \phi)^2\right].
\end{eqnarray}
The sound velocity and Luttinger parameter can be obtained from the Bethe Ansatz solution of the $XXZ$ chain without Zeeman field \cite{giamarchi_2003}
\begin{eqnarray}
\label{Eq:BAS}
\frac{J_z}{J_{xy}} &=& - \cos( \pi \beta^2 ), \nonumber \\
\frac{1}{K} &=& 2 \beta^2,\; v = \frac{1}{1-\beta^2} \sin\left[ \pi (1-\beta^2) \right]\frac{J_{xy}}{2}.
\end{eqnarray}
For our representation of the ladder Hamiltonian it is useful to retain the bosonized forms of the spin operators, whose continuum versions are given by
\begin{equation}
S^{x,y} ( x = j a) = \frac{S_j^{x,y}}{\sqrt{a}}, \; S^z( x = j a ) = \frac{S^z}{a}. 
\end{equation}
These are expressed in terms of the pair $\theta(x),\phi(x)$ as follows:
\begin{equation}
S_z(x) = -\frac{1}{\pi} \nabla \phi(x) + \frac{1}{\pi a} \cos\left[ 2 \phi(x) - 2 k_F x \right],
\end{equation}
where $k_F = \frac{\pi}{2 a}$ for half-filling of the fermions (Jordan-Wigner transformation), equivalent to the $S_z^\text{total} = 0$ sector of the spin Hamiltonian. Turning to the spin ladder operator, one could use either of the following two expressions. If picking a non-Hermitean string operator,
\begin{equation}
S^+(x) = \frac{e^{-i\theta(x)}}{\sqrt{2\pi a}} (-1)^x \left[ 1 + e^{- 2i \phi(x) + 2 i k_F x} \right].
\end{equation}
If we pick a Hermitean string operator $\frac{1}{2}(e^{i  \pi \sum_{j<i} c^\dagger_j c_j  } + \text{H.c.}) = \cos( \sum_{j<i} c^\dagger_j c_j ) $, then 
\begin{equation}
S^+ = \frac{e^{-i \theta(x)}}{\sqrt{2 \pi a}} \left[ (-1)^x + \cos( 2 \phi - 2 k_F x) \right].
\end{equation}
\subsection{Ladder Hamiltonian for hard core bosons}
We now turn to the ladder Hamiltonian in Eq.~(\ref{Eq:H}). The equivalent spin--$1/2$ Hamiltonian is
\begin{eqnarray}
  H &=& -t \sum_{\alpha,i} e^{i a A_{i,i+1}^\alpha} S_{\alpha,i}^+ S_{\alpha,i+1}^- + \text{H.c.}
  \nonumber \\
    && -g \sum_{i} e^{-i a' A_{\perp,i}} S_{1,i}^+ S_{2,i}^- + \text{H.c.} 
  \nonumber \\
    && + V_\perp \sum_i \left( S_{1,i}^z + \frac{1}{2} \right)\left( S_{2,i}^z + \frac{1}{2}\right).
\end{eqnarray}
The bosonized Hamiltonian is expressed in terms of two pairs of fields, each pair defined for one of the chains $\alpha=1,2$, obeying the algebra $[\phi^\alpha(x),\theta^\beta(x')] = i \frac{\pi}{2} \delta^{\alpha\beta} \text{Sign}(x'-x)$. It takes the form
\begin{equation}
H = H_{0}^1 + H_0^2 + H_\perp^g + H_\perp^{V_\perp}.
\end{equation}
In what follows, we introduce $A_\parallel^\alpha(x)$ and $A_{\perp}(x)$, which are continuum equivalents of the lattice quantities $A_{i,i+1}^\alpha$ and $A_{\perp,i}$.

\begin{widetext}
The first two terms are Luttinger liquid Hamiltonians for each spin chain,
\begin{equation}
H_0^\alpha = \frac{v}{2\pi} \int dx \left[ K(\nabla \theta^\alpha - A_\parallel^\alpha)^2 + \frac{1}{K} (\nabla \phi^\alpha)^2  \right].
\end{equation}
The Luttinger parameter and the sound velocity are deducible from Eq.~(\ref{Eq:BAS}). We turn to the coupling Hamiltonian
\begin{equation}
H_\perp^g = -\frac{g}{\pi a} \int dx \cos(-\theta^1 + \theta^2 - a'A_\perp)
\left[  \cos(\pi x) + \cos( 2 k_F^1 x - 2 \phi^1 )  \right]\left[  \cos(\pi x) + \cos( 2 k_F^2 x - 2 \phi^2 )  \right],
\end{equation}
where we replaced $(-1)^x = \cos(\pi x)$. Moreover,
\begin{equation}
H_\perp^{V_\perp} = V_\perp a \int dx \left( \frac{1}{2a} -\frac{\nabla \phi^1}{\pi} + \frac{\cos( 2\phi^1 - 2 k_F^1 x)}{\pi a} \right) \left( \frac{1}{2a} -\frac{\nabla \phi^2}{\pi} + \frac{\cos( 2\phi^2 - 2 k_F^2 x)}{\pi a} \right).
\end{equation}
We recast the quadratic part of the Hamiltonian in a rotated basis in terms of $\phi^\pm = \frac{\phi^1 \pm \phi^2}{\sqrt{2}}$ and $\theta^\pm = \frac{\theta^1 \pm \theta^2}{\sqrt{2}}$, which are still canonically conjugate $[\phi^\alpha(x),\theta^\beta(x')] = i \frac{\pi}{2} \delta^{\alpha\beta} \text{Sign}(x'-x)$ with $\alpha, \beta = \pm$. The quadratic part reads
\begin{eqnarray}
H_0^- &=& \frac{v^-}{2\pi} \int dx \left[ K^- (\nabla \theta^- - A_\parallel^-) + \frac{1}{K^-} (\nabla \phi^-)^2  \right], \nonumber \\
H_0^+ &=& \frac{v^+}{2\pi} \int dx \left[ K^+ (\nabla \theta^+ ) + \frac{1}{K^+} (\nabla \phi^+)^2  \right], \nonumber \\
K^\pm &=& K \left(  1 \pm \frac{V_\perp K a}{\pi v}  \right)^{-1/2}, \nonumber \\
v^{\pm} &=& v \left(  1 \pm \frac{V_\perp K a}{\pi v}  \right)^{1/2}.
\end{eqnarray}
In the rotated basis, the sine-Gordon terms read
\begin{eqnarray}
H_\perp^g &=& -\frac{g}{\pi a} \int dx \cos( - \sqrt{2} \theta^- - a' A_\perp) \times
\nonumber \\
&&\;\;\;\;\;\;\;\;\;\;\left\{ 1 + \cos( 2 k_F^1 x - 2 \phi^1) \cos( 2 k_F^2 x - 2 \phi^2) + \cos\left( \pi x \right) \left[ \cos( 2 k_F^1 x - 2 \phi^1 ) + \cos( 2 k_F^2 x - 2 \phi^2 )  \right]  \right\}.
\end{eqnarray}
For general $k_F^{1,2} \in \left[\frac{\pi}{2a}, \frac{\pi}{a} \right]$, regard the $\cos(\pi x)[...]$  contribution as oscillatory and discard. In addition, let us select the gauge $A_\parallel = 0$, and $a'A_\perp(x) = \chi x$. Using $\cos a \cos b = \frac{1}{2} \left[ \cos(a-b) + \cos(a+b)  \right]$, we obtain
\begin{eqnarray}
H_\perp^g &=& -\frac{g}{\pi a} \int dx \cos( - \sqrt{2} \theta^- - a' A_\perp) \left\{ 1 + \frac{1}{2}\cos\left[ 2 (k_F^1 + k_F^2)x - 2\sqrt{2} \phi^+\right] + \frac{1}{2}\cos\left[ 2 (k_F^1 - k_F^2)x - 2\sqrt{2} \phi^-\right]\right\} 
\nonumber \\
         &&+ \text{ oscillatory terms}.
\end{eqnarray}
Note that the scaling dimension of the second sine-Gordon term in $\{...\}$ is $\frac{1}{2K^-} + 2K^- > 2$, so just keep
\begin{equation}
H_\perp^g = -\frac{g}{\pi a} \int dx \cos( - \sqrt{2} \theta^- - a' A_\perp) \left\{ 1 + \frac{1}{2}\cos\left[ 2 (k_F^1 + k_F^2)x - 2\sqrt{2} \phi^+\right]\right\}.
\end{equation}
Note that if
\begin{equation}
2(k_F^1 + k_F^2) x \pm \chi x = 0 \text{ mod } 2\pi,
\end{equation}
then the contribution $\cos( -\sqrt{2} \theta^- \pm 2 \sqrt{2} \phi^+ )$ is free of oscillatory arguments. Its scaling dimension $\frac{1}{2K^-} + 2 K^+$ will be analyzed below. For the moment, let us write the remainder of the Hamiltonian coming from rung repulsive interactions
\begin{eqnarray}
H_\perp^{V_\perp} = \frac{V_\perp}{2 a \pi^2} \int dx \left\{ \cos\left[ 2 (k_F^1 - k_F^2) - 2\sqrt{2} \phi^- \right] +  \cos\left[ 2 (k_F^1 + k_F^2) + 2\sqrt{2} \phi^+ \right]  \right\} \nonumber \\
+ \text{ terms of the form } \nabla \phi^1 \cos( 2 k_F^2 x - 2 \phi^2 ) \text{ which we drop.}
\end{eqnarray}
For the particular case of hard core bosons without intrachain long range repulsive interactions ($V_\parallel = 0$), $K=1$ so therefore $u = J_{xy} = 2 t$. Therefore the scaling dimensions must be evaluated using
\begin{equation}
K^\pm = \left(  1 \pm \frac{V_\perp}{2\pi t}  \right)^{-1/2}. 
\end{equation}
Note that the Luttinger parameter depends only on the ratio of $V_\perp$ to the bandwidth $\propto t$. Returning to the term $\cos( -\sqrt{2} \theta^- \pm 2 \sqrt{2} \phi^+ )$, we find its scaling dimension
\begin{equation}
\delta = \frac{1}{2 K^-} + 2K^+ = \frac{1}{2} \left( 1 - \frac{V_\perp}{2 \pi t} \right)^{1/2} + 2 \left( 1 + \frac{V_\perp}{2 \pi t} \right)^{-1/2}.
\end{equation}
Numerically, we find that 
\begin{equation}
\delta < 2 \text{ for } \frac{V_\perp}{t} \geq \pi,
\end{equation}
so interactions which are sizeable compared to the tunneling rate are necessary to make the Laughlin sine--Gordon term relevant.  
\end{widetext}

\section{Edge theory in Laughlin phase}
\label{App:ET}
In this appendix we derive the effective edge Hamiltonian in Luttinger liquid form. We make use of the Fourier transform conventions of App.~\ref{App:FTC}. The outline of our strategy is as follows: consider the gapped $c=1$ Hamiltonian in the Laughlin phase. Assume that the interchain coupling flows to infinity $g\to \infty$. Expand the sine--Gordon term to quadratic order to obtain a mass term. Integrate out the gapped field and its canonically conjugate phase variable,
to obtain the gapless edge Hamiltonian, defined by Luttinger parameter $K^e$ and sound velocity $v^e$.

Our starting point is the Hamiltonian~(\ref{Eq:HWB}) composed of Luttinger liquid parts for the ``+'' sector~(\ref{Eq:LLp}), the ``-'' sector~(\ref{Eq:LLm}), and the Laughlin sine--Gordon term (\ref{Eq:LaughlinCouplingm}).  The corresponding action is
\begin{eqnarray}
  \label{Eq:MinusS}
  -S  &=& -\frac{1}{\Omega} \sum_{\alpha=\pm,q} 
  \left(
  \begin{array}{cc}
    \theta_q^{\alpha *} & \phi_q^{\alpha *}
  \end{array}
  \right)
  \left( 
  \begin{array}{cc}
    \frac{v^\alpha K^\alpha}{2\pi} k^2 & \frac{i k \omega_n}{2\pi} \\
    \frac{i k \omega_n}{2\pi} & \frac{v^\alpha}{2\pi K^\alpha} k^2 
  \end{array}
  \right)
  \left( 
  \begin{array}{c}
    \theta^\alpha_q \\
    \phi^\alpha_q
  \end{array}
  \right) \nonumber \\
  && - 2 g n_0 \int dr \cos( \sqrt{2} \theta^- - m \sqrt{2}\phi^+ ),
\end{eqnarray}
with $\Omega = \beta L$ (see App.~\ref{App:FTC} for notations). The action $S$ is obtained from the Hamiltonian via a Legendre transform \cite{giamarchi_2003}. The Gaussian part is expressed in the momentum representation, whereas the nonquadratic part requires an expansion to quadratic order, to be performed below.

This action is expressed in terms of the original fields $\theta^\pm$ and $\phi^\pm$. 

A rotation to a basis in which the gapped field $\sqrt{2}\theta^- - m \sqrt{2} \phi^+$ is isolated is desired. We introduce chiral fields \cite{kane_et_al_2002}
\begin{equation}
  \label{Eq:Chirals}
  \phi_r^\alpha = \frac{\theta^\alpha}{m} + r \phi^\alpha,
\end{equation}
where $r=+$ for right moving fields and $r=-$ for left moving fields, and $\alpha=1,2$ is the chain index. These fields obey the algebra
\begin{equation}
  \label{Eq:ChiAlgebra}
  [\phi_r^\alpha(x), \phi^\beta_{p}(x')] = i r \delta_{\alpha \beta} \delta_{rp} \frac{\pi}{m} \text{sgn}(x'-x).
\end{equation}
The above is equivalent to Eq.~(\ref{Eq:A}) via the relation between phase--density fields and chiral fields~(\ref{Eq:Chirals}).

Define
\begin{eqnarray}
  \phi &=& \left( - \phi_{-1}^{1} + \phi_{+1}^2 \right) / 2, \nonumber \\
  \theta &=& \left( + \phi_{-1}^{1} + \phi_{+1}^{2} \right) / 2, \nonumber \\
  \phi'  &=& \left( - \phi_{-1}^{2} + \phi_{+1}^{1} \right) / 2, \nonumber \\
  \theta' &=& \left( + \phi_{-1}^2 + \phi_{+1}^{1} \right) / 2. \label{Eq:PhiChi}
\end{eqnarray}
In terms of the original fields,
\begin{eqnarray}
  \phi &=&  -\frac{\theta^-}{m\sqrt{2}} + \frac{\phi^+}{\sqrt{2}}, \nonumber \\
  \theta &=&  \frac{\theta^+}{m\sqrt{2}} - \frac{\phi^-}{\sqrt{2}}, \nonumber \\
  \phi' &=&  \frac{\theta^-}{m\sqrt{2}} + \frac{\phi^+}{\sqrt{2}}, \nonumber \\
  \theta' &=&  \frac{\theta^+}{m\sqrt{2}} + \frac{\phi^-}{\sqrt{2}}. \label{Eq:PhiBasis}
\end{eqnarray}
Using the above set of relations we obtain commutators:
\begin{equation}
  \label{Eq:PhiThetaCommutator}
  [\phi(x),m\theta(x')] =  [\phi'(x),m\theta'(x') ] = i \frac{\pi}{2} \text{sgn}(x'-x),
\end{equation}
whereas
\begin{equation}
  [\phi(x),m\theta'(x')] = [\phi'(x),m\theta(x')] = 0,
\end{equation}
where the prime on the field and the prime on the coordinate are unrelated. Note that $m \theta'(x)$ and $m \theta(x)$ are required in order to satisfy the canonical commutator~\ref{Eq:A}.

The field $e^{-i\theta(x)}$ creates a quasiparticle of charge $1/m$ at $x$. Identifying the charge operator $n(x) = \frac{-1}{\pi}\nabla_x \phi(x)$, we find that $\theta(x)$ produces a kink in $\phi(x)$ that corresponds to the addition of a fractional charge $1/m$
\begin{eqnarray}
  e^{i \theta(x')} n(x) e^{-i\theta(x')} &=& n (x) - \frac{1}{2m} \nabla_x \text{sgn}(x'-x)  \nonumber \\
  &=& n(x) + \frac{1}{m} \delta(x-x').
\end{eqnarray}
That is, $e^{-i\theta(x')}$ creates a particle of charge $1/m$ at position $x'$. Consistently, $n(x)$ is the boson charge operator. A quasiparticle at $x'$ corresponds to $1/m \delta(x-x')$ added to this field. The considerations of this paragraph hold equally for the doublet $\theta',\phi'$.

The chiral boson creation operators for the edge are
\begin{equation}
  e^{- i m \phi_{+1}^{1}} , e^{ - i m \phi_{-1}^2},
\end{equation}
where an edge boson is created by operator
\begin{equation}
\psi^{e\dagger}(x) =  e^{- i m \theta'} + e^{i m\pi n_0^e x} e^{- i m (\theta' + \phi')} + e^{-i m\pi n_0^e x} e^{ - i m (\theta' - \phi')},
\end{equation}
which qualifies $m\theta'(x)$ as the bosonic phase field of the edge theory. The fields $m\theta',\phi'$ are the aimed-for Luttinger liquid fields for the gapless edge theory. 

The chiral boson creation operators for the bulk Hamiltonian are
\begin{equation}
  e^{- i m \phi_{-1}^{1}} , e^{ - i m \phi_{+1}^2}.
\end{equation}
The sine--Gordon term responsible for the Laughlin gap may be reexpressed as $- 2 g n_0 \int dr \cos( 2 m \phi ) = - 2 g n_0 \int dr \cos m \left( - \phi_{-1}^{1} + \phi_{+1}^2 \right)$. It is the gap inducing backscattering of the bulk chiral fields \cite{kane_et_al_2002}. The fields $\theta$ and $\phi$ are to be integrated.

For a Gaussian integration to be possible, we expand the sine--Gordon term on the second row of the action~(\ref{Eq:MinusS}) 
\begin{eqnarray}
  &&- 2 g n_0 \int dr \cos(2 m \phi)  \nonumber \\ &=& - 2 g n_0 \Omega + 2 g n_0 \frac{1}{2} 4 m^2  \int dr [\phi(r)]^2 + O(\phi^4) \nonumber \\
  &=& + 4 g n_0 m^2 \frac{1}{\Omega} \sum_q \phi_q \phi_q^* - 2 g n_0 \Omega + O(\phi^4).
\end{eqnarray}

\begin{widetext}
Moreover,
\begin{equation}
  \label{Eq:DefR}
  \left( 
  \begin{array}{c}
    \theta^+ \\
    \phi^+ \\
    \theta^- \\
    \phi^-
  \end{array}
  \right)
  =
  \left( 
  \begin{array}{cccc}
    \frac{m}{\sqrt{2}} & 0 & \frac{m}{\sqrt{2}} & 0 \\
    0 & \frac{1}{\sqrt{2}} & 0 & \frac{1}{\sqrt{2}} \\
    0 & \frac{-m}{\sqrt{2}} & 0 & \frac{m}{\sqrt{2}} \\    
    \frac{-1}{\sqrt{2}} & 0 & \frac{1}{\sqrt{2}} & 0     
  \end{array}
  \right)
  \left(
  \begin{array}{c}
    \theta \\
    \phi \\
    \theta' \\
    \phi'
  \end{array}
  \right) \equiv R    \left(
  \begin{array}{c}
    \theta \\
    \phi \\
    \theta' \\
    \phi'
  \end{array}
  \right),
\end{equation}
such that
\begin{eqnarray}
  - S = - \sum_q 
  \left(
  \begin{array}{cccc}
    \theta_q^* & \phi_q^* & \theta_q^{'*} & \phi_q^{'*} 
  \end{array}
  \right)
  M
  \left(
  \begin{array}{c}
    \theta_q \\ \phi_q \\ \theta_q^{'} \\ \phi_q^{'} 
  \end{array}
  \right).
\end{eqnarray}
with
\begin{eqnarray}
  M &\equiv&  \frac{1}{\Omega}   R^\dagger 
  \left( 
  \begin{array}{cccc}
    \frac{v^+ K^+}{2\pi} k^2 & \frac{i k \omega_n}{2\pi} & 0 & 0 \\
    \frac{i k \omega_n}{2\pi} & \frac{v^+}{2\pi K^+} k^2 & 0 & 0 \\
    0 & 0 & \frac{v^- K^-}{2\pi} k^2 & \frac{i k \omega_n}{2\pi} \\
    0 & 0 & \frac{i k \omega_n}{2\pi} & \frac{v^-}{2\pi K^-} k^2 
  \end{array}
  \right) 
  R
  \nonumber 
  -  \frac{1}{\Omega} 
  \left( 
  \begin{array}{cccc}
    0 & 0 & 0 & 0 \\
    0 & 4 g n_0 m^2 & 0 & 0 \\
    0 & 0 & 0 & 0 \\
    0 & 0 & 0 & 0     
  \end{array}
  \right).
\end{eqnarray}
The matrix $M$ is symmetric with complex $q$--dependent entries.  

Gaussian integration of fields $\theta,\phi$ gives new quadratic contributions in the $\theta', \phi'$ due to off diagonal terms in $M$. The resulting effective edge action is
\begin{equation}
- S(\theta',\phi')= - \sum_q 
\left( 
\begin{array}{cc}
  \theta^{'*}_q & \phi^{'*}_q 
\end{array}
\right)
  \left(
  \begin{array}{cc}
    P_{\theta',\theta'} & P_{\theta',\phi'} \\
    P_{\phi',\theta'} & P_{\phi',\phi'} 
  \end{array}
  \right) \left( 
\begin{array}{c}
  \theta^{'}_q \\ \phi^{'}_q 
\end{array}
\right),
\end{equation}
where the entries of the matrix are given by
  \begin{eqnarray}
  P_{\theta',\theta'} &=&  \frac{\left(K^+ v^+ m^2+\frac{v^-}{K^-}\right) k^2}{4 \pi  \Omega }-\frac{\left(m^2 K^+v^+-\frac{v^-}{K^-}\right)^2 k^2}{4 \pi  \Omega  \left(K^+ v^+ m^2+\frac{v^-}{K^-}\right)} \nonumber \\
  &&+\frac{m^2 \omega _n^2 \left(m^2 K^+ v^+-\frac{v^-}{K^-}\right)^2 k^2}{4 \pi ^2 \Omega  \left(K^+ v^+ m^2 +\frac{v^-}{K^-}\right)^2 \left(\frac{\left(K^- v^- m^2+\frac{v^+}{K^+}\right) k^2}{4 \pi }-4 g m^2 n_0 +\frac{m^2 \omega _n^2}{\pi  \left(K^+ v^+ m^2+\frac{v^-}{K^-}\right)}\right)},  \nonumber \\
  P_{\theta',\phi'} &=& \frac{i m \omega _n \left(\frac{v^+}{K^+}-m^2 K^-
   v^-\right) \left(m^2 K^+ v^+-\frac{v^-}{K^-}\right) k^3}{8 \pi ^2 \Omega  \left(K^+ v^+
   m^2+\frac{v^-}{K^-}\right) \left(-4 g n_0 m^2+\frac{\omega _n^2 m^2}{\pi  \left(K^+ v^+
   m^2+\frac{v^-}{K^-}\right)}+\frac{m^2 K^- v^- k^2+\frac{v^+ k^2}{K^+}}{4 \pi
   }\right)}+\frac{i m \omega _n k}{2 \pi  \Omega },    \nonumber \\
  P_{\phi',\theta'} &=& \frac{i m \omega _n \left(\frac{v^+}{K^+}-m^2 K^- v^-\right) \left(m^2 K^+
   v^+-\frac{v^-}{K^-}\right) k^3}{8 \pi ^2 \Omega  \left(K^+ v^+ m^2+\frac{v^-}{K^-}\right)
   \left(\frac{\left(K^- v^- m^2+\frac{v^+}{K^+}\right) k^2}{4 \pi }-4 g m^2 n_0+\frac{m^2
   \omega _n^2}{\pi  \left(K^+ v^+ m^2+\frac{v^-}{K^-}\right)}\right)}+\frac{i m \omega _n
   k}{2 \pi  \Omega },    \nonumber \\
  P_{\phi',\phi'}   &=& \frac{k^2 \left(K^- v^- m^2+\frac{v^+}{K^+}\right)}{4 \pi  \Omega
  }-\frac{k^4 \left(\frac{v^+}{K^+}-m^2 K^- v^-\right)^2}{16 \pi ^2 \Omega 
    \left(\frac{\left(K^- v^- m^2+\frac{v^+}{K^+}\right) k^2}{4 \pi }-4 g m^2 n_0+\frac{m^2
      \omega _n^2}{\pi  \left(K^+ v^+ m^2+\frac{v^-}{K^-}\right)}\right)}.    \nonumber
  \end{eqnarray}
An expansion of $P$ in $1/g$ shows that every additional power of $\frac{1}{g}$ multiplies either $\omega_n^2$ or $k^2$. Keeping only second order derivatives amounts to putting $g\to\infty$ in the above
\begin{eqnarray}
  P_{\theta',\theta'} &\approx&  \frac{\left(K^+ v^+ m^2+\frac{v^-}{K^-}\right) k^2}{4 \pi  \Omega }-\frac{\left(m^2 K^+ v^+-\frac{v^-}{K^-}\right)^2 k^2}{4 \pi  \Omega  \left(K^+ v^+ m^2+\frac{v^-}{K^-}\right)}, \nonumber \\
  P_{\theta',\phi'} &\approx& \frac{i k m \omega _n}{2 \pi  \Omega }, \nonumber \\
  P_{\phi',\theta'} &\approx& \frac{i k m \omega_n}{2 \pi  \Omega } ,  \nonumber \\
  P_{\phi',\phi'}   &\approx& \frac{k^2 \left(K^- v^- m^2+\frac{v^+}{K^+}\right)}{4 \pi  \Omega }.\nonumber
\end{eqnarray}
\end{widetext}

Rescale the phase field, such that $\theta' \to m\theta'$, leading to $[\phi'(x),\theta'(x')]= i \frac{\pi}{2} \text{sgn}(x'-x)$. The following adjustments of the coefficients above reflect this change:
\begin{eqnarray}
  P_{\theta',\theta'} &\approx&  \frac{\left(K^+ v^+ +\frac{v^-}{m^2 K^-}\right) k^2}{4 \pi  \Omega }-\frac{\left( K^+ v^+-\frac{v^-}{m^2 K^-}\right)^2 k^2}{4 \pi  \Omega  \left(K^+ v^+ +\frac{v^-}{m^2 K^-}\right)}, \nonumber \\
  P_{\theta',\phi'} &\approx& \frac{i k \omega _n}{2 \pi  \Omega }, \nonumber \\
  P_{\phi',\theta'} &\approx& \frac{i k \omega_n}{2 \pi  \Omega } ,  \nonumber \\
  P_{\phi',\phi'}   &\approx& \frac{k^2 \left(K^- v^- m^2 +\frac{v^+}{K^+}\right)}{4 \pi  \Omega }.\label{Eq:RescaleThetaP}
\end{eqnarray}
To identify the coefficients of the generic Luttinger liquid action~\cite{giamarchi_2003}, in terms of ``edge'' Luttinger parameter and velocity $K^e,v^e$, we require that
\begin{eqnarray}
  P_{\theta',\theta'} &=& \frac{1}{\Omega} \frac{v^{e} K^{e}}{2\pi}k^2, \nonumber \\
  P_{\theta',\phi'} &=& \frac{1}{\Omega} \frac{i k \omega_n}{2\pi} = P_{\phi',\theta'}, \nonumber \\
  P_{\phi',\phi'} &=& \frac{1}{\Omega} \frac{v^e}{2\pi K^e}k^2.
\end{eqnarray}

Note that the middle condition, corresponding to the Berry phase term, is  already satisfied. The other two conditions amount to
\begin{eqnarray}
  \label{Eq:veKe}
  v^e K^e &=& \frac{K^+ v^+ +\frac{v^-}{m^2 K^-}}{2} - \frac{\left( K^+ v^+ -\frac{v^-}{m^2 K^-}\right)^2 }{2 \left(K^+ v^+ +\frac{v^-}{m^2 K^-}\right)}, \\
  \label{Eq:veoKe}
  v^e / K^e &=& \frac{K^- v^- m^2 +\frac{v^+}{K^+}}{2}.
\end{eqnarray} 
Using the form for $K^\pm$ and $v^\pm$ in Eq.~(\ref{Eq:vKWB}), we have, letting $h^{\pm} = 1 \pm \frac{V_\perp K a}{\pi v}$: 
\begin{eqnarray}
  v^e K^e &=& \frac{vK + h^- \frac{v}{m^2 K}}{2} 
  - \frac{\left( v K - h^- \frac{v}{m^2 K} \right)^2}{2\left( v K + h^- \frac{v}{m^2 K} \right)} \nonumber \\ 
  &\to& v\left[ \frac{ K + \frac{1}{m^2 K}}{2} - \frac{\left( K - \frac{1}{m^2 K} \right)^2}{2\left( K + \frac{1}{ m^2 K} \right)} \right],
\end{eqnarray}
the latter if  $V_\perp \to 0$, and
\begin{equation}
  \frac{v^e}{K^e} = v \frac{ m^2 K + h^+ \frac{1}{K}}{2} \to v \frac{ m^2 K + \frac{1}{K}}{2},
\end{equation}
the latter if  $V_\perp \to 0$. Calculating $K^e$ for $V_\perp = 0$ and $K = 1$ and $m=2$ (hard core bosons, and the most relevant Laughlin term corresponding to picking the lowest density harmonic at $m=2$) gives
\begin{equation}
  K^e = 2/5 = 0.4,
\end{equation}
while the sound velocity remains unmodified
\begin{equation}
  v^e = \frac{2}{5} \frac{2^2 + 1}{2} v = \frac{2}{5} \frac{5}{2} v  = v.
\end{equation}

We have arrived at the edge Hamiltonian
\begin{equation}
  \mathcal{H}^e = \frac{v}{2\pi} \int_0^L dx \left[ K^e(\nabla \theta')^2 + \frac{1}{K^e} (\nabla \phi')^2 \right].
\end{equation}
It is instructive to recast this in terms of the edge chiral fields $\phi^{1}_{+1}(x) \equiv R(x)$ and $\phi_{-1}^2 \equiv L(x)$ which leads to
\begin{eqnarray}
  \mathcal{H}^e &=& \frac{v}{2\pi} \int_0^L dx \left[ K^e(\nabla \theta')^2 + \frac{1}{K^e} (\nabla \phi')^2 \right]  \\
  &=& \frac{v}{8\pi} \int_0^L dx \left[ m^2 K^e(\nabla R-\nabla L)^2 + \frac{1}{K^e} (\nabla R + \nabla L)^2 \right] \nonumber \\
  &=& \frac{v}{8\pi} \int_0^L dx \left[ A_{RR} (\nabla R)^2 + A_{LL} (\nabla L)^2 + A_{LR}(\nabla R)(\nabla L)  \right], \nonumber \\
\end{eqnarray}
where we have accounted for the rescaling of the field $\theta'$, and have introduced coefficients
\begin{eqnarray}
  A_{RR} &=& A_{LL} =  m^2 K^e + \frac{1}{K^e}, \nonumber \\
  A_{LR} &=& \frac{2}{K^e} - 2 m^2 K^e.
\end{eqnarray}
There is backscattering between the edge chiral fields, resulting from the traced bulk degrees of freedom.  

Remark that an edge theory without backscattering would mandate long range repulsive interactions. Setting $K=1/m$ at $V_\perp=0$, one would have
\begin{eqnarray}
  v^e K^e = \frac{v}{m}, \; v^e / K^e = v m,
\end{eqnarray}
which makes $K^e = 1/m$ and $v^e = v$, and $A_{LR}=0$, while $A_{LL}=2m = A_{RR}$ at $m=2$, i.e.
\begin{equation}
  \mathcal{H}^e = \frac{mv}{4\pi} \int_0^L dx \left[ (\nabla R)^2 + (\nabla L)^2 \right].
\end{equation}
We have suppressed backscattering terms in the edge theory of the two chain model by long range interactions and have obtained an edge theory identical to the chiral Luttinger liquid discussed by Wen \cite{wen_1995,*wen_1992}.

\section{Fourier transform conventions}
\label{App:FTC}
This section summarizes our conventions for Fourier transforms in App.~\ref{App:ET}. Fields are assumed periodic on the interval $[0,L]$, where $L$ is the system length. The Fourier decomposition of a field is
\begin{equation}
  f(r) = \frac{1}{\Omega} \sum_q f_q e^{iqr},
\end{equation}
where $\Omega = \beta L$, $\beta = 1/T$, $r = (x,v \tau)$, $q = (k,\omega_n / v)$, and $qr = kx -\omega_n \tau$, and $\omega_n = \frac{2\pi n}{\beta}$ for $n$ integer are the Matsubara frequencies. This makes $\sum_q \equiv \sum_{n\text{ integer}} \sum_k$ in the above equation. The Fourier transform of a field is
\begin{equation}
  f_q = \int dr f(r) e^{-iqr},
\end{equation}
where $\int dr \equiv \int_0^\beta d\tau \int_0^L dx$. The two equations above allow us to write the resolution of the $\delta-$function
\begin{eqnarray}
  \delta_{q,q'} &=& \frac{1}{\Omega} \int dr e^{-i(q-q')r}, \nonumber \\
  \delta^{(2)}(r-r') &=& \frac{1}{\Omega} \sum_{q} e^{i q(r-r')}. 
\end{eqnarray}
Using these formulae, we arrive at
\begin{eqnarray}
  \int dr f(r)g(r) &=& \frac{1}{\Omega} \sum_q f_q g_{-q} \nonumber \\ 
  &=& \frac{1}{\Omega} \sum_q f_q g_q^*, \nonumber \\
  \int dr \partial_\tau f(r) g(r) &=& \frac{1}{\Omega} \sum_q f_q (-i\omega_n) g_{-q} \nonumber \\ &=& \frac{1}{\Omega} \sum_q f_q (-i\omega_n) g_{q}^*, \nonumber \\
  \int dr \nabla f(r) g(r) &=& \frac{1}{\Omega} \sum_q f_q (+ i k )g_{-q} \nonumber \\
  &=& \frac{1}{\Omega} \sum_q f_q (+ i k )g^*_{q},
\end{eqnarray}
where each second row holds if $g(r)$ is real for all $r$.

\section{Fits for central charge}
\label{App:CC}
In this appendix, we present a more precise method to obtain the central charge as the coefficient of the logarithmic contribution in the entanglement entropy.

\begin{figure}[t!]
  a)\includegraphics[width=0.95\linewidth]{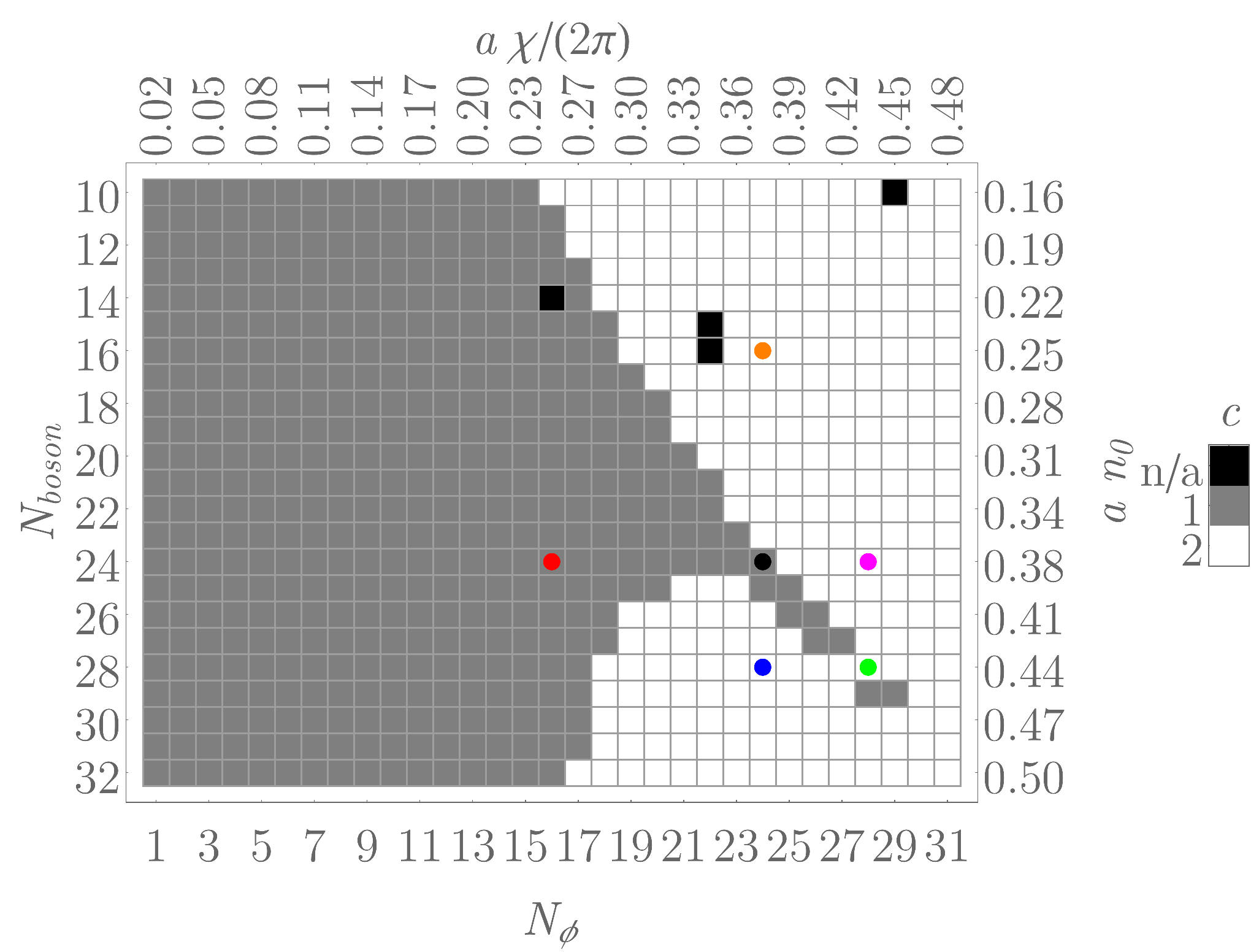}
  b)\includegraphics[width=0.95\linewidth]{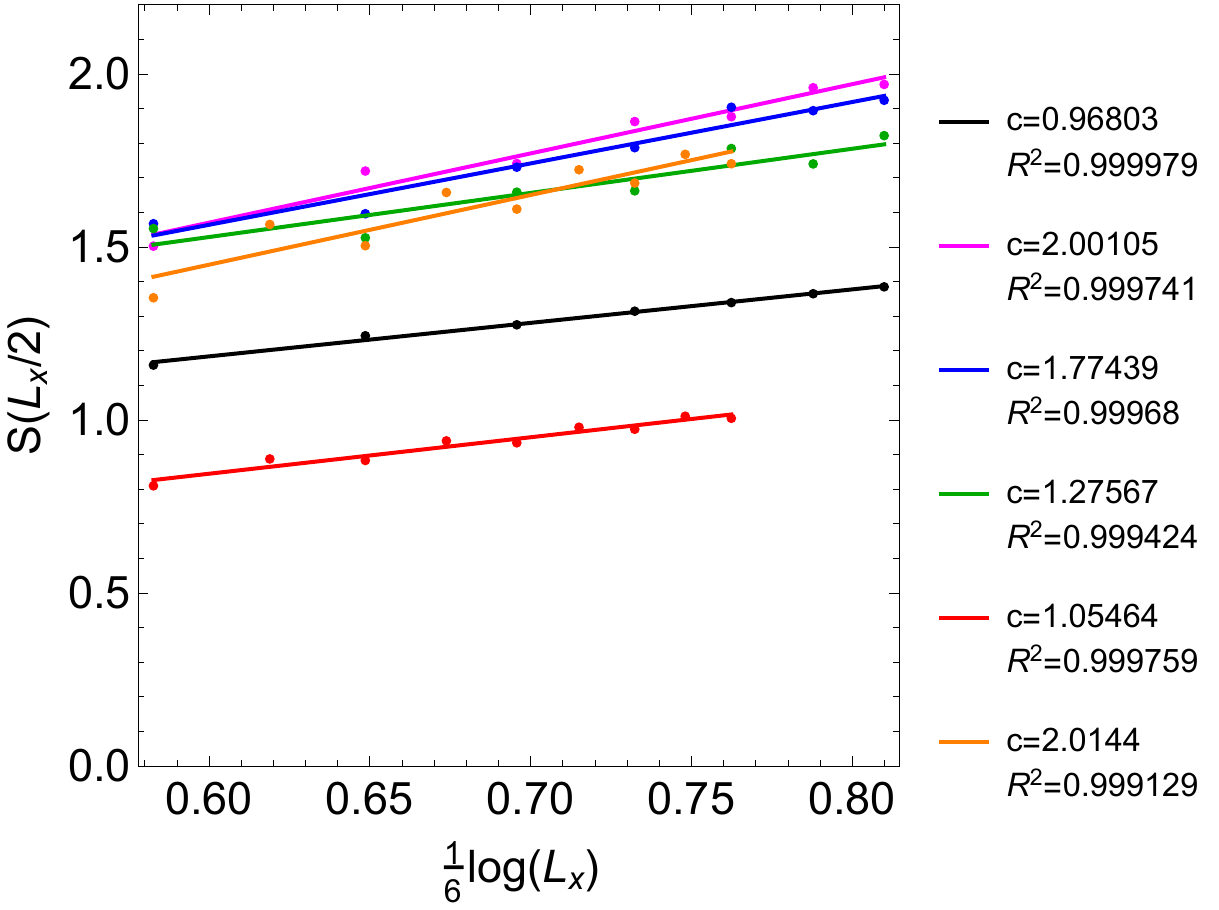}
  \caption{\label{Fig:Ap:CC}(Color online) Central charge from fits of the bipartite entanglement entropy, obtained by finite size scaling (Panel b)) from~(\ref{Eq:Ap:CCL}). Colors correspond to $(n_0,\chi)$ pairs marked in a), which contains fits at on a 65 rung ladder according to Eq.~(\ref{Eq:SF}) with $c$ constrained to the set $\{1,2\}$.}
\end{figure}

Figure~\ref{Fig:Ap:CC}a) contains the results of fits for the central charge according to  Eq.~(\ref{Eq:SF}) of the main text, while constraining $c=0,1$ or $2$ and retaining the fit with the best $R^2$. The system size is $L=65$ with open boundary conditions. Values of the entropy corresponding to 4 sites at each end of the system were discarded. 

Alternatively, one can express using Eq.~(\ref{Eq:SF}) the bipartite entanglement entropy evaluated at the middle bond:
\begin{equation}
  \label{Eq:Ap:CCL}
  S\left[ (L+1)/2 \right] = \frac{c}{6} \log\left( L \right) + B,
\end{equation}
where $B$ is the bond energy~\cite{laflorencie_et_al_2006,roux_et_al_2009,cardy_calabrese_2010} of Eq.~(\ref{Eq:SF}) which becomes independent of $L$ for large enough $L$. Then $c$ arises from a linear fit, in which the system length can be progressively increased. The results obtained for points marked with colored asterisks on Fig.~\ref{Fig:Ap:CC}a) are shown in Fig.~\ref{Fig:Ap:CC}b), with acceptable convergence throughout the phase diagram.

\section{Ground state degeneracy for thin torus}
\label{App:GSD}
This appendix contains details for the derivations in Sec.~\ref{Subsec:GSD}.  

We can find the relation between the eigenvectors $| \psi_\alpha (2\pi) \rangle$ and $| \psi_\alpha ( 0 ) \rangle$, for $\alpha = 0,1$, as follows. We have
\begin{eqnarray}
H(2\pi) |\psi_\alpha(2\pi) \rangle = E_\alpha( 2\pi ) | \psi_\alpha ( 2\pi ) \rangle \nonumber \\ = U_y^\dagger( 2\pi ) H( 0 ) U_y( 2\pi ) | \psi_\alpha ( 2\pi ) \rangle.
\end{eqnarray}
The first line is a definition in the text,~Eq.~(\ref{Eq:EigenSysNot}), and the second is the unitary transformation relating $H(0)$ and $H(2\pi)$, Eq.~(\ref{Eq:H_gtransform}). Applying $U_y(2\pi)$ to the left, we get
\begin{equation}
H(0) U_y(2\pi) | \psi_\alpha(2\pi) \rangle  = E_\alpha( 2\pi ) U_\alpha( 2\pi ) | \psi_\alpha( 2\pi ) \rangle.
\end{equation}
This means that the vectors $U_y(2\pi) |\psi_\alpha(2\pi) \rangle$ for $\alpha = 1,2$ are the eigenstates of $H(0)$ with eigenvalues $E_\alpha(2\pi)$. Further assuming that these eigenvalues are nondegenerate, that is $E_0(0) = E_0(2\pi) < E_1(0) = E_1(2\pi)$, we obtain 
\begin{equation}
|\psi_\alpha (0) \rangle = U_y(2\pi) |\psi_\alpha(2\pi) \rangle.
\end{equation}

How can the eigenvectors $|\psi_\alpha(2\pi) \rangle$ be related to the eigenvectors $|\psi_\alpha(0) \rangle$? For this, inspect the eigenvalues of the momentum operator:
\begin{equation}
\label{AppEq:PyEE}
P_y U_y(2\pi) | \psi_\alpha ( 2\pi ) \rangle = U_y( 2\pi ) U_y^\dagger( 2\pi ) P_y U_y( 2\pi ) | \psi_\alpha ( 2\pi ) \rangle.
\end{equation}
To evaluate the eigenvalue in the above, we need to know the action of the gauge transformation on the momentum operator:
\begin{eqnarray}
&&U_y^\dagger( 2\pi ) P_y U_y( 2\pi ) \nonumber \\ &=& \frac{1}{L_y} \sum_{k_y} \sum_{i,j,j'} U_y^\dagger(2\pi) b^\dagger_{j,i} b_{j',i} U_y(2\pi)  e^{i k_y ( j - j')} k_y  \nonumber \\
 &=& \frac{1}{L_y} \sum_{k_y} \sum_{i,j,j'} b_{j,i}^\dagger b_{j',i} e^{i \pi( j-j' )} e^{i k_y (j-j')} k_y \nonumber \\
 &=& \frac{1}{L_y} \sum_{k_y} \sum_{i,j,j'} b_{j,i}^\dagger b_{j',i} e^{i(k_y + \pi) (j-j')}( k_y + \pi - \pi) \nonumber \\
 &=& \frac{1}{L_y} \sum_{\tilde{k_y} \equiv k_y + \pi} \sum_{i,j,j'} b_{j,i}^\dagger b_{j',i} e^{i \tilde{k_y} (j-j')} \tilde{k_y} \nonumber\\ &&\;\;\;\;- \pi \frac{1}{L_y}\sum_{k_y} \sum_{i,j,j'} b_{j,i}^\dagger b_{j',i} e^{i k_y (j-j')} \nonumber \\
 &=& P_y - \pi \sum_{j,i} b^\dagger_{j,i} b_{j,i} 
= P_y - \pi N.
\end{eqnarray}
In the derivation above, we have set $L_y = 2$ and used that in its sum $k_y$ takes the values $\{ 0, \pi\}$ in units of the inverse lattice constant. Returning then to the eigenvalue equation~(\ref{AppEq:PyEE}), we find
\begin{equation}
\label{AppEq:PyDifference}
P_y U_y(2\pi) |\psi_\alpha (2\pi )\rangle = \left[  P_{y\alpha}(2\pi) - \pi N  \right] U_y( 2\pi ) | \psi_\alpha( 2\pi ) \rangle.
\end{equation}
If there is an odd number of particles, for example $N=3$ as considered for exact diagonalization in the text, this means that $P_{y\alpha}(2\pi) - 3\pi = P_{y,\alpha}(0)$, meaning that the two states $|\psi_\alpha(0) \rangle$ and $|\psi_\alpha(2\pi)\rangle$ lie in distinct momentum sectors.

Using~(\ref{Eq:EnergyEquality}), we can conclude that if the two lowest lying energy states of $H(0)$ [and consequently the two lowest lying states of $H(2\pi)$] are nondegenerate then, since $E_{\alpha}(0) = E_{\alpha}(2\pi)$, 
\begin{equation}
|\psi_\alpha (0) \rangle = U_y(2\pi) |\psi_\alpha(2\pi) \rangle.
\end{equation}
However, the corresponding momentum quantum numbers differ as expressed in~(\ref{AppEq:PyDifference}) and this difference is dependent on the number of particles. In particular, note that for $N=3$ particles $E_0(\theta_y)$ and $E_1(\theta_y)$ must touch at least once for $\theta_y$ in the interval $[ 0, 2\pi )$. For, assuming there is always a finite gap between these two states, then $|\psi_\alpha(\theta_y) \rangle$ must remain an eigenstate of $P_y$ with eigenvalue $P_{y\alpha}(0)$, for all values of $\theta_y$, but this is impossible as $P_{y\alpha}(2\pi) = P_{y\alpha}(0)+\pi$ for 3 particles. However, for $N=2,4$, such a level crossing is unnecessary.

\bibliography{references}
\bibliographystyle{unsrtnat}
\bibliographystyle{apsrev4-1}

\end{document}